\documentstyle[psfig]{mn}

\newif\ifAMStwofonts

\ifoldfss
  \ifCUPmtlplainloaded \else
    \NewTextAlphabet{textbfit} {cmbxti10} {}
    \NewTextAlphabet{textbfss} {cmssbx10} {}
    \NewMathAlphabet{mathbfit} {cmbxti10} {} 
    \NewMathAlphabet{mathbfss} {cmssbx10} {} 
  \fi
  \ifAMStwofonts
    \ifCUPmtlplainloaded \else
      \NewSymbolFont{upmath} {eurm10}
      \NewSymbolFont{AMSa} {msam10}
      \NewMathSymbol{\upi}     {0}{upmath}{19}
      \NewMathSymbol{\umu}     {0}{upmath}{16}
      \NewMathSymbol{\upartial}{0}{upmath}{40}
      \NewMathSymbol{\leqslant}{3}{AMSa}{36}
      \NewMathSymbol{\geqslant}{3}{AMSa}{3E}

      \let\leq=\leqslant \let\le=\leqslant
       \let\ge=\geqslant
    \fi
  \fi
\fi 

\ifnfssone
  \newmathalphabet{\mathit}
  \addtoversion{normal}{\mathit}{cmr}{m}{it}
  \addtoversion{bold}{\mathit}{cmr}{bx}{it}
  \newmathalphabet{\mathbfit} 
  \addtoversion{normal}{\mathbfit}{cmr}{bx}{it}
  \addtoversion{bold}{\mathbfit}{cmr}{bx}{it}
  \newmathalphabet{\mathbfss} 
  \addtoversion{normal}{\mathbfss}{cmss}{bx}{n}
  \addtoversion{bold}{\mathbfss}{cmss}{bx}{n}
  \ifAMStwofonts
    \ifCUPmtlplainloaded \else
      \UseAMStwoboldmath
      \makeatletter
      \new@mathgroup\upmath@group
      \define@mathgroup\mv@normal\upmath@group{eur}{m}{n}
      \define@mathgroup\mv@bold\upmath@group{eur}{b}{n}
      \edef\UPM{\hexnumber\upmath@group}
      \new@mathgroup\amsa@group
      \define@mathgroup\mv@normal\amsa@group{msa}{m}{n}
      \define@mathgroup\mv@bold\amsa@group{msa}{m}{n}
      \edef\AMSa{\hexnumber\amsa@group}
      \makeatother
      \mathchardef\upi="0\UPM19
      \mathchardef\umu="0\UPM16
      \mathchardef\upartial="0\UPM40
      \mathchardef\leqslant="3\AMSa36
      \mathchardef\geqslant="3\AMSa3E

      \let\leq=\leqslant \let\le=\leqslant
       \let\ge=\geqslant
    \fi
  \fi
\fi 

\ifnfsstwo
  \DeclareMathAlphabet{\mathbfit}{OT1}{cmr}{bx}{it}
  \SetMathAlphabet\mathbfit{bold}{OT1}{cmr}{bx}{it}
  \DeclareMathAlphabet{\mathbfss}{OT1}{cmss}{bx}{n}
  \SetMathAlphabet\mathbfss{bold}{OT1}{cmss}{bx}{n}
  \ifAMStwofonts
    \ifCUPmtlplainloaded \else
      \DeclareSymbolFont{UPM}{U}{eur}{m}{n}
      \SetSymbolFont{UPM}{bold}{U}{eur}{b}{n}
      \DeclareSymbolFont{AMSa}{U}{msa}{m}{n}
      \DeclareMathSymbol{\upi}{0}{UPM}{"19}
      \DeclareMathSymbol{\umu}{0}{UPM}{"16}
      \DeclareMathSymbol{\upartial}{0}{UPM}{"40}
      \DeclareMathSymbol{\leqslant}{3}{AMSa}{"36}
      \DeclareMathSymbol{\geqslant}{3}{AMSa}{"3E}

      \let\leq=\leqslant \let\le=\leqslant
       \let\ge=\geqslant
    \fi
  \fi
\fi 

\ifCUPmtlplainloaded \else
  \ifAMStwofonts \else 
    \def\upi{\pi}
    \def\umu{\mu}
    \def\upartial{\partial}
  \fi
\fi

\title[Origin of E+A's]{Origin of E+A galaxies: I. Physical properties of E+A's
formed from galaxy  merging and interaction}
\author[K. Bekki,  W.  J. Couch, Y. Shioya, A. Vazdekis]
       {K. Bekki,${}^1$W. J. Couch${}^1$   Y. Shioya${}^2$, and A. Vazdekis${}^3$ \\
        ${}^1$School of Physics, University of New South Wales, Sydney 2052, NSW, Australia \\
        ${}^2$Astronomical Institute, Tohoku University, Sendai, 980-8578, Japan\\
        ${}^3$ Instituto de Astrofisica de Canarias, La Laguna 38200, Tenerife, Spain}
\date{Accepted 
      Received
      in original form 2001}

\pagerange{\pageref{firstpage}--\pageref{lastpage}}
\pubyear{1994}

\begin{document}

\maketitle

\label{firstpage}

\begin{abstract}

We investigate the structural, kinematical, and spectrophotometric 
properties of ``E+A'' galaxies -- those  
with strong Balmer absorption lines but no significant [OII] emission -- 
using numerical simulations combined with stellar population synthesis codes. 
We particularly focus on the two-dimensional (2D) distributions of
line-of-sight velocity, velocity dispersion, colors, line index
in E+A galaxies formed via the interaction and merging 
of two gas-rich spirals.
Our numerical simulations demonstrate that
E+A elliptical galaxies formed by major galaxy merging 
have  positive radial color gradients
and negative radial H$\delta$ gradients by virtue of  
their central poststarburst populations. Furthermore, we show that the
projected kinematical and spectroscopic properties of the simulated 
E+A galaxies can be remarkably different for different major merger models.
For example, the simulated E+A ellipticals with kinematically decoupled cores
clearly show regions of strong H$\delta$ absorption which are very 
flattened, with differences in rotation and velocity dispersion between
the old and young stars. E+A ellipticals are highly likely to show more 
rapid rotation and a smaller central velocity dispersion
in young stars than in old ones.
E+A's formed from the strong tidal interaction between gas-rich spirals
have disky morphologies with thick disks and are highly likely to be morphologically
classified as barred S0s.  
We also provide specific predictions on the
structural, kinematical, and spectrophotometric properties
of young globular cluster systems in E+A's. 
Based on these results, we discuss the importance of spatially resolved, integral 
field unit spectroscopy on large (8-10m) ground-based telescopes 
in confirming the formation of kinematically distinct cores
in elliptical galaxies produced via dissipative merging  
and determining the most probable physical mechanism(s)
for E+A formation with disky and spheroidal morphologies.
\end{abstract}


\section{Introduction}

Since Dressler \& Gunn (1983) discovered the enigmatic "E+A" galaxies -- i.e., those 
with strong Balmer line absorption but no detectable emission --  in distant clusters, 
their origin and the physical mechanisms responsible for their formation have long been
discussed both observationally and theoretically. There have been two key issues which
have driven much of this discussion: Firstly, what is responsible for triggering and then 
abruptly truncating the
recent starburst that most likely gave rise to the E+A spectral signature (Poggianti 
et al. 1999)? Secondly, what is the evolutionary link between E+A galaxies and the 
prevalent 
blue 'Butcher-Oemler' (Butcher \& Oemler 1978) population? Although the nature of 
E+A galaxies in distant clusters has been extensively studied observationally 
(Couch \& Sharples 1987, Barger et al. 1996, Couch et al. 1998, Dressler et al. 1999, 
Galaz 2000, Tran et al. 2003), these two issues remain far from being resolved.

Another important avenue of investigation has been to study the incidence of E+A galaxies 
in environments {\it other} than distant clusters, mainly at lower redshifts using large 
galaxy redshift surveys. Zabludoff et al. (1996, hereafter Z96) investigated 21 
low-redshift (0.05 $<$ $z$ $<$ 0.13) E+A's identified in a sample of 11,113 galaxies 
observed as part of the Las Campanas 
Redshift Survey. They found that most of these E+A galaxies lay in the field rather 
than in clusters and therefore suggested that cluster environmental effects, such as 
interactions with the cluster gravitational potential or intracluster
medium, are not responsible for E+A formation. More recently, Blake et al. (2004) 
conducted a similar investigation using the order of magnitude larger 2dF Galaxy
Redshift Survey (2dFGRS), and found the large majority of the $\sim$50--250 $z\sim 0.1$ 
E+A galaxies in their sample were either isolated or in galaxy groups, with only 
$\sim 10\%$ being in rich clusters. At higher redshifts, Tran et al. (2004) investigated 
the fraction of E+A galaxies in a sample of $\sim$ 800 spectroscopically confirmed field 
galaxies in the range $0.3\leq z\leq 1.0$ and found it to be $\sim$ 2.7 $\pm$ 1.1\%, 
which is significantly lower than that in galaxy clusters at comparable redshifts 
(11 $\pm$ 3\%).

Recent high resolution imaging, photometric and spectroscopic studies of E+A
galaxies undertaken with the $Hubble$ $Space$ $Telescope$ ($HST$) and large 
ground--based telescopes have revealed a considerable diversity in the morphological, 
structural, and kinematical properties between E+A's.
Norton et al. (2001) investigated the internal kinematics of the E+A galaxies
in the Z96 sample and found that most of them were dynamically
supported, with $v/\sigma$  and $\sigma$ ranging from 30\,km\,s$^{-1}$ to 
200\,km\,s$^{-1}$ (here $v$ is the rotational velocity and $\sigma$ is the velocity
dispersion). A very recent HST-based morphological study of 5 of the Z96 E+A's by 
Yang et al. (2004), revealed that they were  
bulge-dominated systems (with and without a small underlying disk)
and had radial luminosity profiles qualitatively similar to those of normal 
``power-law'' early-type galaxies. They also discovered possible young and
very bright ($M_{\rm R}$ $<$ $-13$) globular cluster candidates around these
E+A galaxies, with the number varying significantly from galaxy to galaxy. 
Tran et al. (2004) have also revealed the diversity in 
half-light radius, total luminosity, morphological type, and internal velocity dispersion
between E+A's in their  samples. However, Blake et al. (2004) have cautioned that
there maybe some heterogeneity and hence diversity in E+A galaxy samples due to the
way they are selected; E+A samples whose strong Balmer line absorption has been identified 
only the basis of the equivalent width measured for the H$\delta$ (and sometimes
the H$\gamma$) line -- as has generally been the case in distant cluster studies 
(Dressler et al. 1999) -- will be contaminated by a population of disk-dominated, dusty
star-forming galaxies, as well as containing {\it bona fide} E+A's which are 
spheroid-dominated and have no ongoing star formation.

This plethora of observational data on E+A galaxies, when taken in its entirety, 
raise many questions, the most significant being: (i)\,Can model(s) of galaxy 
interactions and merging (e.g., minor vs major merging) explain {\it self-consistently}
the observed dynamical, photometric, and spectroscopic properties of E+A's?.
(ii)\,What mechanism drives the formation of disky E+A's that dominates E+A populations 
in clusters of galaxies? (iii)\,Are there any evolutionary links between the  
disky E+A's in clusters and ``passive spirals'' with k-type spectra (Couch et al. 1998)? 
(iv)\,What is the origin of the positive and negative color gradients observed in E+A's?
(v)\,Which can more self-consistently explain the very red colors
of E+A's discovered both in the field and in clusters -- dust extinction
or a truncated IMF? (vi)\,Are there any differences between the structural and kinematical
properties of the old and young stellar populations in E+A galaxies?  
(vii)\,Are the physical properties of bright young star clusters
observed around E+A's consistent with any formation scenario of
globular cluster formation? Although previous theoretical studies have tried
to understand possible star formation histories of E+A galaxies
(Barbaro \& Poggianti 1997; Poggianti et al. 1999; Shioya \& Bekki 2000;
Shioya et al. 2002; 2004), they had difficulty in addressing the above questions
due to the limitations of the one-zone spectrophotometric models that they adopted.
In order to properly tackle the above questions, the dynamical and 
spectrophotometric properties of E+A galaxies need to be investigated 
{\it jointly} in an explicitly self-consistent manner.
Therefore, numerical simulations combined with stellar
population synthesis codes -- which enable us to predict not only structural and 
kinematical properties but also photometric and spectroscopic ones -- 
are indispensable in solving the above important problems related to
the formation and evolution of E+A galaxies.

The purpose of this paper is to investigate the structural, 
kinematical, chemical,  photometric, 
and spectroscopic properties of E+A galaxies in an explicitly self-consistent
manner, thereby making considerable progress towards answering the above 
question. In particular, we investigate both {\it radial gradients} in the 
spectrophotometric properties as well as the {\it projected} (2D) distributions of 
Balmer absorption, rotational velocity and velocity dispersion in E+A galaxies formed
from galaxy interactions and merging:
We will discuss other possible mechanisms of E+A galaxy formation in our
forthcoming papers.
 These spatial properties are the main focus
of our study since current and planned spatially resolved spectroscopy with 
integral field units (IFUs) on 8-10m telescopes (e.g., GMOS on Gemini)
is going to provide an unprecedented wealth of data on the two-dimensional
distribution of spectroscopic properties for E+A's (Pracey et al. 2004). 
Furthermore observational studies based on the SAURON project
with reasonably large field of views  
(e.g., Davies et al. 2001) now  provide 2D spectra data
of kinematics and  stellar populations for the entire regions of galaxies.  
The simulated 2D distributions of the dynamical and spectroscopic properties 
will not only enable us to identify peculiar structural and kinematical behaviour
(e.g., kinematically decoupled cores) associated with the young stars in E+A galaxies,   
but also help us to coverage on the most reasonable E+A formation models. 
Our models can also predict the age-, metallicity- and spatial-distributions of
young GCs associated with E+A galaxies, and so we can discuss in more depth 
the bright young candidates recently observed by Yang et al. (2004).

The plan of this paper is as follows: In the next section,
we briefly describe our numerical models for E+A formation via 
galaxy interactions and merging. We summarize the numerical methods and
techniques by which we can calculate the spectrophotometric
properties of E+A's from N-body numerical data. 
In \S 3, we present the numerical results on the physical properties of E+A's
for different merger/interaction models.
In \S 4, we use our results to address each of the seven questions mentioned
above. We summarize our conclusions in \S 5.

In presenting our work here, we stress that we are focussing solely on
those E+A galaxies formed as a result of galaxy-galaxy interactions
and merging. This does not, however, rule out other mechanisms
(e.g., abrupt truncation of star formation via halo gas stripping and
ram pressure stripping) being responsible for the formation of
E+A galaxies, and that different mechanisms can operate in
different environments (see, for example, Barbaro \& Poggianti 1997;
Bekki et al. 2001a,b; Bekki et al. 2002).

\begin{table}
\centering
\begin{minipage}{185mm}
\caption{Model parameters}
\begin{tabular}{cccccl}
model no. & 
{$f_{\rm b}$ %
\footnote{mass ratio of bulge to disk}} 
& {$f_{\rm g}$ %
\footnote{gas mass fraction}}
& {$m_2$  %
\footnote{mass ratio of two merging spirals}}
& orbit 
& comments \\ 
M1 & 0.5 &  0.1 & 1.0 & PP &  major merger\\
M2 & 0.5 &  0.2 & 1.0 & PP &  \\
M3 & 0.5 &  0.5 & 1.0 & PP &  \\
M4 & 0.0 &  0.1 & 1.0 & IN &  no bulge\\
M5 & 0.5 &  0.1 & 1.0 & IN &  fiducial \\
M6 & 0.5 &  0.2 & 1.0 & IN &   \\
M7 & 0.5 &  0.5 & 1.0 & IN &   \\
M8 & 1.0 &  0.1 & 1.0  & IN &   more massive bulge\\
M9 & 0.0 &  0.1 & 1.0 & IN &   no bulge\\
M10 & 0.5 &  0.1 & 1.0 & RR &   \\
U1 & 0.5 &  0.1 & 0.3 & IN &  unequal-mass merger \\
U2 & 0.5 &  0.1 & 0.1 & IN &  \\
T1 & 0.0 &  0.1 & 1.0 & PP &  tidal interaction \\
T2 & 0.0 &  0.2 & 1.0 & PP &   \\
T3 & 0.0 &  0.5 & 1.0 & PP &   \\
T4 & 0.0 &  0.5 & 1.0 & RR &   \\
T6 & 1.0 &  0.1 & 1.0  & PP &   \\
T7 & 0.0 &  0.1 & 0.1  & PP &   \\
I1 & 0.0 &  0.1 & -  & - &  isolated disk \\
I2 & 0.5 &  0.1 & - & - &   \\
I3 & 1.0 &  0.1 & - & - &   \\
\end{tabular}
\end{minipage}
\end{table}

\section{The model}

\subsection{Mergers}

Since the numerical methods and techniques we employ for modeling the chemodynamical
and photometric evolution of galaxy mergers have already been described in detail 
elsewhere (Bekki \& Shioya 1998, 1999), we give only  a brief review here. 

\subsubsection{Progenitor disk galaxies}

The progenitor disk galaxies that take part in a merger are taken to 
have a dark halo, a bulge, and a thin exponential disk.
Their total mass and size are $M_{\rm d}$ and $R_{\rm d}$, respectively. 
From now on, all masses are measured in units of
$M_{\rm d}$ and  distances in units of $R_{\rm d}$, unless otherwise specified. 
Velocity and time are measured in units of $v$ = $ (GM_{\rm d}/R_{\rm d})^{1/2}$ and
$t_{\rm dyn}$ = $(R_{\rm d}^{3}/GM_{\rm d})^{1/2}$, respectively,
where $G$ is the gravitational constant and assumed to be 1.0
in the present study. 
If we adopt $M_{\rm d}$ = 6.0 $\times$ $10^{10}$ $ \rm M_{\odot}$ and
$R_{\rm d}$ = 17.5\,kpc as fiducial values, then $v$ = 1.21 $\times$
$10^{2}$\,km\,s$^{-1}$  and  $t_{\rm dyn}$ = 1.41 $\times$ $10^{8}$ yr.

We adopt the density distribution of the NFW 
halo (Navarro, Frenk \& White 1996) suggested from CDM simulations:
 \begin{equation}
 {\rho}(r)=\frac{\rho_{0}}{(r/r_{\rm s})(1+r/r_{\rm s})^2},
 \end{equation} 
 where  $r$, $\rho_{0}$, and $r_{\rm s}$ are
the spherical radius,  the central density of a dark halo,  and the scale
length of the halo, respectively.  
The value of $r_{\rm s}$ (0.8) is chosen such that
the rotation curve of the disk is reasonably consistent with
observations. The bulge has a density profile
with a shallow cusp (Hernquist 1990): 
 \begin{equation}
 {\rho}(r) \propto r^{-1}(r+a_{\rm bulge})^{-3},
 \end{equation} 
where $a_{\rm bulge}$ is the scale length of the bulge
and fixed at 0.04.
The bulge mass and its compactness can control the bar formation
in the disks and thus the strength of starbursts in mergers. 
The dark matter to disk mass ratio 
is fixed at 9 whereas the bulge to disk ratio 
is assumed to be a free parameter and represented by $f_{\rm b}$.
The radial ($R$) and vertical ($Z$) density profiles 
of the  disk are  assumed to be
proportional to $\exp (-R/R_{0}) $ with scale length $R_{0}$ = 0.2
and to  ${\rm sech}^2 (Z/Z_{0})$ with scale length $Z_{0}$ = 0.04
in our units, respectively.
In addition to the rotational velocity attributable to the gravitational
field of the disk and halo components, the initial radial and azimuthal velocity
dispersions are added to the disk component in accordance with
the epicyclic theory, and with a Toomre parameter value of $Q$ = 1.5
(Binney \& Tremaine 1987) .
The vertical velocity dispersion at a given radius 
is set to be 0.5 times as large as
the radial velocity dispersion at that point, 
as is consistent with the trend observed in the Milky Way (e.g., Wielen 1977).

The disk is composed both of gas and stars, with the gas mass fraction
($f_{\rm g}$) being a free parameter and the gas disk
represented by a collection of discrete gas clouds that follow the observed mass-size
relationship (Larson 1981). All overlapping pairs of gas clouds
are made to collide with the same restitution coefficient of 0.5
(Hausman \& Roberts 1984). The gas is converted into either {\it field stars} 
or {\it globular clusters (GCs)}. Field star formation
is modeled by converting  the collisional gas particles
into  collisionless new stellar particles according to the algorithm
of star formation  described below. We adopt the Schmidt law (Schmidt 1959)
with exponent $\gamma$ = 1.5 (1.0  $ < $  $\gamma$
$ < $ 2.0, Kennicutt 1998) as the controlling
parameter of the rate of star formation. The amount of gas 
consumed by star formation for each gas particle
in each time step is given by:
\begin{equation}
\dot{{\rho}_{\rm g}} \propto  
{\rho_{\rm g}}^{\gamma},
\end{equation}
where $\rho_{\rm g}$ 
is the gas density around each gas particle. 
The coefficients in the law are taken from the work of Bekki (1998, 1999) 
and the mean star formation rate in an isolated disk model
for 1 Gyr evolution is $\sim$ 1 ${\rm M}_{\odot}$ for the adopted
coefficient (thus consistent with observations). 
Globular cluster formation in the present model is discriminated from
field star formation as follows.
We use the cluster formation criteria derived by
previous analytical works (e.g., Kumai et al. 1993), 
and hydrodynamical simulations  with a variety of different parameters for
cloud-cloud collisions on a 1-100\,pc scale (Bekki et al. 2004) 
in order to model globular
cluster formation. A gas particle is converted into a cluster if it
collides with other high velocity gas (with the relative velocities
ranging from 30\,km\,s$^{-1}$
to 100\,km\,s$^{-1}$) and having an impact parameter (normalized to the
cloud radius) less than 0.25. 
These stars formed from gas are called ``new stars'' (or ``young stars'')
whereas stars initially within a disk are called ``old stars''. 
throughout this paper.

\subsubsection{Chemical and spectrophotometric evolution}

Chemical enrichment through star formation during galaxy merging
is assumed to proceed both locally and instantaneously in the present study.
We assign the metallicity of the original
gas particles to  the new stellar particles and increase 
the metallicity of the neighbouring gas particles. 
The total number of neighbouring gas particles is taken to be $N_{\rm gas}$,
given by the following equation for chemical enrichment:
  \begin{equation}
  \Delta M_{\rm Z} = \{ Z_{i}R_{\rm met}m_{\rm s}+(1.0-R_{\rm met})
 (1.0-Z_{i})m_{\rm s}y_{\rm met} \}/N_{\rm gas}. 
  \end{equation}
Here, $\Delta M_{\rm Z}$ represents the increase in metallicity for each
gas particle, $ Z_{i}$ the metallicity of the new stellar particle (or that
of the original gas particle), $R_{\rm met}$ the fraction of gas returned
to the interstellar medium, $m_{\rm s}$ the mass of the new star,
and $y_{\rm met}$ the chemical yield.
The values of $R_{\rm met}$ and $y_{\rm met}$ are set to 0.3 and 0.02, respectively.
It is assumed here that the spectral energy distribution (SED) of a model galaxy is 
the sum of the SEDs of the individual stellar particles. 
The SED of each  stellar particle is assumed to be  
a simple stellar population (SSP) that is  
a coeval and chemically homogeneous assembly of stars. 
Thus the monochromatic flux of a galaxy with age $T$,
$F_{\lambda}(T)$,  is described as 
\begin{equation}
F_{\lambda}(T) = \sum_{\rm star} F_{{\rm SSP},\lambda}(Z_{i},
{\tau}_{i}) \times m_{\rm s} \; ,
\end{equation}
where $F_{{\rm SSP},\lambda}(Z_{i},{\tau}_{i})$ and $m_{\rm s}$ 
are the monochromatic flux of a SSP 
of age ${\tau}_{i}$ and metallicity $Z_{i}$ (where the suffix $i$ identifies 
each stellar particle),  and the 
mass of each stellar particle, respectively.
The age of each SSP, ${\tau}_{i}$, is defined as ${\tau}_{i} = T - t_{i}$, 
where $t_{i}$ is the time when a gas particle is converted into a stellar one.
The metallicity of each SSP is exactly the same
as that  of the stellar particle, $Z_{i}$, and the summation ($\sum$) in
equation (5) is done  for all 
stellar particles in a model galaxy.

The adopted instantaneous recycling approximation  means that 
the model does not consider the time delay between star formation
and the onset of supernovae explosion. This means that 
although chemical enrichment associated with type II SNe 
(that form  $\sim$ $10^7$ yr after star formation)
can be modeled reasonably well,
the one  associated with SNeIa 
(that form  $\sim$ $10^9$ yr after star formation)
is not so correctly modeled 
owing to the large time delay between star formation and SNIa explosion. 
As a result of this, the present model can not correctly predict each chemical
abundance (e.g., Mg and Fe) in E+A's: We have a plan to investigate
chemical properties of E+A's in our forthcoming papers by
using more sophisticated  models including SNIa effects.

A stellar particle is assumed to be composed of stars whose
age and metallicity are exactly the same as those of the stellar particle
and the total mass of the stars is set to be the same as that of
the  stellar particle.
Thus the monochromatic flux of a SSP at a given wavelength is defined as
\begin{equation}
F_{{\rm SSP}, \lambda}(Z_{i},{\tau}_{i}) = \int_{M_L}^{M_U} 
\phi (M) f_{\lambda}(M, {\tau}_{i}, Z_{i}) dM \; ,
\end{equation}
where $M$ is the mass of the star, and $f_{\lambda}(M, {\tau}_{i}, Z_{i})$
is the monochromatic flux of a star with mass $M$, metallicity $Z_{i}$ and age 
${\tau}_{i}$. $\phi (M)$ is the initial mass function (IMF) of stars and 
$M_U$, $M_L$ are the upper and lower mass limits of the IMF, respectively. 
In this paper, we use the $F_{{\rm SSP}, \lambda}(Z_{i}, {\tau}_{i})$ values
calculated by Vazdekis et al. (1996) with an IMF slope of 1.3
(i.e., the Salpeter IMF).
We also adopt their two different models with $M_L=0.1\,{\rm M}_{\odot}$ and 
1.0\,${\rm M}_{\odot}$.

In order to calculate the spectrophotometric evolution of merger remnants
with poststarburst components, we need to assume that 
(1)\,the initial stellar disk and bulge in a merger progenitor spiral 
have the same age of $T_{\rm disk}$,  
and (2)\,the disk has a metallicity gradient consistent with observations.
Considering the recent observations of the Galactic metallicity gradient (Friel 1995),
we allocate metallicity to each disc star according to its initial position:
At $r$ = $R$, where $r$($R$) is the projected distance (in units of kpc)
from the center of the disc, the metallicity of the star is given by:
\begin{equation}
{\rm [m/H]}_{\rm r=R} = {\rm [m/H]}_{\rm d, r=0} + {\alpha}_{\rm d} \times {\rm R}. \;
\end{equation}
We adopt a  plausible value of $-0.091$ for the slope ${\alpha}_{\rm d}$  
from Friel (1995), in which the Galactic stellar metallicity gradient
is estimated from Galactic open clusters.
The central value of ${\rm [m/H]}_{\rm d, r=0}$ is chosen such that
${\rm [m/H]}_{\rm r=R}$ is 0.02 at the radius of 8.5 kpc (the solar radius
for the Galactic disk).
The merger models with the adopted metallicity gradients above are 
demonstrated to explain consistently the observed metallicity
distribution function of the stellar halo in NGC 5128 (Bekki et al. 2003).
We investigate the models with $T_{\rm disk}$ = 5, 7, 10\,Gyr, because
our results can be compared with the relatively nearby ($z$ $\sim$ $0.1-0.3$)
E+A's. We show only the results  with $T_{\rm disk}=7$\,Gyr in the present study,
because the results do not strongly depend on $T_{\rm disk}$ for the above range
of $T_{\rm disk}$. 
Several authors have suggested that starbursts produce a top-heavy IMF,
because the formation of lower-mass stars is suppressed during such an event 
(e.g., Larson 1998). Considering  this possible difference in the IMF between 
starburst phases and periods of normal star formation, 
we adopt $M_{\rm L}$ = 0.1 ${\rm M}_{\odot}$ 
for old stars and 
$M_{\rm L}$ = 1.0 ${\rm M}_{\odot}$ for new stars formed during starbursts
in deriving $F_{\rm SSP}$ for each stellar component.

The main reason for our adopting a 'top-heavy' IMF for the young starburst
component is as follows: Shioya et al. (2004) have demonstrated that
poststarburst models with a normal IMF
(with $M_{\rm L}$  = 0.1 ${\rm M}_{\odot}$) 
are unable to simultaneously explain
the red colors and strong H$\delta$ absorption that are seen for
a subset of the E+A galaxies observed by 
Couch \& Sharples (1987), Balogh et al. (1997),
and Caldwell et al. (1999).
We need
either dust extinction (e.g., Shioya et al. 2004)
or a top-heavy IMF (Balogh et al. 1997; Charlot et al. 1993)
to explain these red-H$\delta$-strong
E+A's where starbursts have ended:  A top-heavy IMF is a possibility
for explaining the origin of these red E+A galaxies. Therefore we have chosen
a top-heavy IMF for the young stars in the present study.
If the normal IMF is adopted for young stars, the poststarburst 
signature becomes rather weak in E+A phases.
 For example,
EW(H$\delta$) in the standard model is 7.4 \AA  \,
for  $M_{\rm L}$  = 1.0 ${\rm M}_{\odot}$
and 1.1 \AA   \,
for $M_{\rm L}$  = 0.1 ${\rm M}_{\odot}$ at the same time $T$.

The present model does not consider the effects of dust extinction
on the SEDs of poststarburst galaxies, nor the effects of 
gaseous emission on the Balmer absorption lines
(It should be stressed here that spectral indices are not affected
by dust emission and Balmer lines might be affected by nebular
emission).
Since most of the gas in an interacting/merging pair 
should be consumed in the associated starburst and hence by the time it develops
an E+A spectral signature, the omission of dust extinction effects is justified.
We nonetheless discuss the possible important effects of dust in E+A spectra
in \S 4. In order to incorporate gaseous emission into our model SEDs, we need to 
combine our N-body simulation code with {\it both} the adopted SSP code
(Vazdekis et al. 1996) and with one that can provide emission lines luminosities 
(e.g., PEGASE by Fioc \& Rocca-Volmerange 1997). To combine these three different 
numerical codes is a formidable task, particularly into our new chemodynamical
model which already includes a variable IMF and globular cluster formation. 
Hence we will leave a full treatment of gaseous emission for a future paper.
We therefore adopt the assumption that the dilution of the Balmer
absorption features by emission from interstellar gas ionized by residual
star formation in the post-starburst phase does not occur at all.
Star formation is therefore assumed to be completely 
halted $\sim$ $1.0$\,Gyr
after the star formation associated with the starburst reaches its maximum rate.
We therefore do not intend to discuss the origin of a small amount of possibly
residual star formation
observed in a few E+A galaxies (Miller \& Owen 2001).
AGN feedback effects from massive black holes on 
spectrophotometric properties of E+A's can be also
important (Shioya \& Bekki 2005).

\subsubsection{Orbital configurations}

In all of the simulations of merging pairs, the orbit of the two disks is set to be
initially in the $xy$ plane and the distance between
the center of mass of the two disks
is  assumed  to be 10 in our units (corresponding to 175\,kpc). 
The pericenter distance and the eccentricity
are set to be 1.0 (17.5\,kpc) and 1.0 (i.e., parabolic), 
respectively, for most of the models.
The spin of each galaxy in a merger
is specified by two angles $\theta_{i}$ and
$\phi_{i}$, where suffix  $i$ is used to identify each galaxy.
$\theta_{i}$ is the angle between the $z$ axis and the vector of
the angular momentum of a disk.
$\phi_{i}$ is the azimuthal angle measured from the $x$ axis to
the projection of the angular momentum vector of a disk onto the $xy$ plane. 
We specifically present the results of the following three 
models with different disk inclinations with respect to the
orbital plane: A prograde-prograde model represented by ``PP''
with $\theta_{1}$ = 0, $\theta_{2}$ = 30, $\phi_{1}$ = 0;
a retrograde-retrograde model (``RR'') with $\theta_{1}$ = 180,
$\theta_{2}$ = 210, $\phi_{1}$ = 0, and $\phi_{2}$ = 0; and
a highly inclined model (``IN'') with $\theta_{1}$ = 60, $\theta_{2}$ =
60, $\phi_{1}$ = 90, and $\phi_{2}$ = 0.
The time taken for the progenitor disks to completely merge and reach 
dynamical equilibrium is less than 16.0 in our units ($\sim$ 2.2\,Gyr) for most of
our major merger models. We also present the results of unequal-mass merger models
with the mass ratios of the two merging spirals (represented by $m_2$)
equal to 0.1 and 0.3.
Table 1 summarises the model parameters for the merger models, with
major merger models and unequal-mass models labeled as ``M'' and ``U'',
respectively.

\begin{figure}
\psfig{file=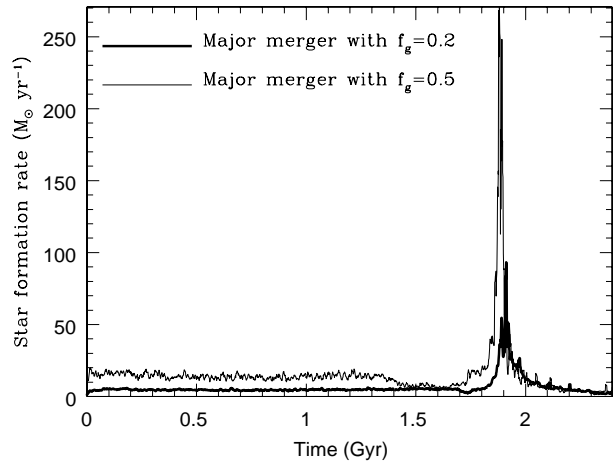,width=8.cm}
\caption{ 
Star formation histories of major galaxy mergers between gas-rich
spirals for the model M2 with  $f_{\rm g}$ = 0.2 ({\it thin solid}) 
and the model M5 with $f_{\rm g}$ =  0.5 ({\it thick solid})
for 2.4\,Gyr. These two models are examples of merger-driven massive starbursts
with a subsequent rapid decline in star formation rate.
}
\label{Figure. 1} this possible difference of IMF between starburst phases
and normal ones in galaxy evolution,
\end{figure}

\begin{figure}
\psfig{file=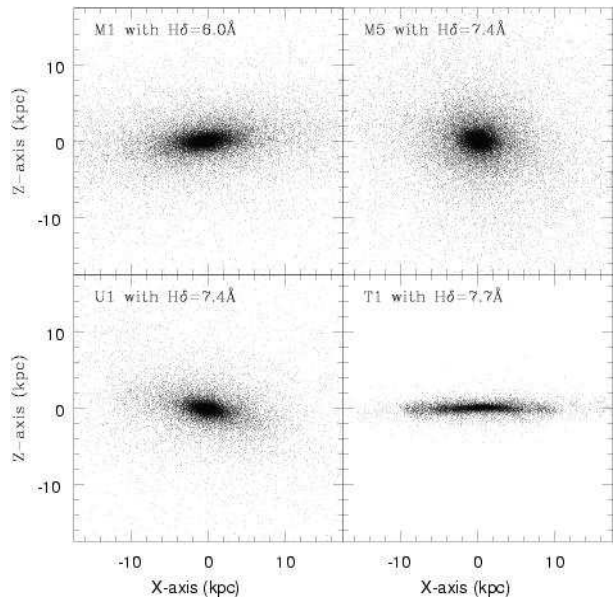,width=8.cm}
\caption{ 
Mass distribution of stars of the simulated E+A's 
projected onto the $x$-$z$ plane
(corresponding to the orbital plane) for four representative 
models: M1 ({\it upper left}), M5 (fiducial, {\it upper right}),
U1 ({\it lower left}), and T1 ({\it lower right}).
EW(H$\delta$) in each model is shown in the upper part of each frame.  
Note that all models show large EW(H$\delta$) ($\ge$ 6 \AA).
}
\label{Figure. 2}
\end{figure}

\subsection{Tidal interaction}

The initial disk models and the numerical methods used for computing  
chemodynamical and spectrophotometric evolution are exactly the same for the
tidal interaction models as they are for the merger models. In the tidal interaction 
models, the two disk galaxies do not merge with each other by virtue of the large
pericenter distance (2 in our units, corresponding to 35\,kpc) that is adopted.
Therefore, only one of the two interacting galaxies needs to be modeled as a fully
self-gravitating N-body system, while the other is modeled as a point-mass particle. 
We mainly present the results of the tidal interaction models with $m_2$ = 1.0:
Strong starbursts do not happen in interaction models with smaller $m_2$ ($\sim$ 0.1); 
hence these models do not have poststarburst populations with strong E+A spectra.
Moderately strong starbursts, with a maximum star formation rate of 
$20-30$ ${\rm M}_{\odot}$, occur in tidal interactions with $m_2$ = 1.0; 
consequently most of the gas within interacting disks is consumed within $\sim 1$\,Gyr.

Table 1 summarises the model parameters for the tidal interaction models, all of 
which are labeled  ``T''.  
For comparison, we also investigated isolated disk models (without tidal interaction
and merging) and these models are labeled as ``I'' in Table 1.
The results of these isolated models are used when we investigate
how strongly star formation rates are increased in interacting/merging galaxies
compared with an isolated disk with no external perturbation.
Star formation rates in these isolated disk models 
are at most a few ${\rm M}_{\odot}$ yr$^{-1}$ whereas
those in interacting/merging models are an order of $10-100$ ${\rm M}_{\odot}$ yr$^{-1}$. 
The results of the isolated models are only very briefly described in
the present study, simply because they do not show E+A spectra.

All the calculations related to the above chemodynamical evolution
have been carried out on the GRAPE board (Sugimoto et al. 1990)
at the Astronomical Data Analysis Center (ADAC)
at the National Astronomical Observatory of Japan. 
The gravitational softening parameter was fixed at 0.025 in our
units (0.44\,kpc). The time integration of
the equation of motion was performed by using the 2nd-order leap-frog method.
The initial total particle number in each simulation was
110,000 for a `merger' model and 55,000 for the `interaction' and 'isolated' models.

\subsection{Main points of analysis}

Star formation histories and morphological evolution of  
interacting and merging galaxies with 
strong starbursts  have already been described in previous numerical studies
(e.g., Noguchi \& Ishibashi 1988; Mihos \& Hernquist 1996; Bekki 1998). 
Furthermore,  spectral type evolution (e.g., from ``e(a)'' to ``a+k''; Dressler et al. 
1999) in starbursting galaxy mergers has been extensively discussed by Bekki et al. (2001a)
based on spectro-dynamical simulations.
In this paper, therefore, we particularly focus on (1)\,the spatial distribution of 
the optical and near-infrared colors and the equivalent widths of the Balmer absorption 
lines in E+A's, and (2)\,the difference in the kinematical properties  of the old and 
young stellar components in E+A galaxies.

As far as (2) is concerned, Norton et al. (2001) found possible differences in 
rotational velocity and velocity dispersion between the old and young components
in nearby E+A's. Our investigation of (2) therefore involves checking  
whether the simulated kinematics of E+A galaxies  
formed by galaxy interactions and merging are consistent with such observations.
The kinematical properties of the young stellar populations in E+A's are derived from
the properties of the strong Balmer absorption lines, which are the key observables
in this context. To compare these observables with the present numerical results
in a consistent way, we show the kinematics of young H($\delta$) strong stars that 
have EW(H$\delta$) larger than 2\AA.  We attempt here to completely remove the 
contribution from new stars that are formed in the very early phase of merging 
(and thus have relatively old ages) and consequently do not contribute to the
strong Balmer line absorption seen in the E+A phase.
 
We also investigate the physical properties of the simulated GC's in E+A's,
because we consider that the predicted properties of young GC's in E+A's
are useful in discussing whether the adopted cloud-cloud collision model
of GC formation is consistent with the observations of young
blue GC candidates in E+A's (e.g., Yang et al. 2004),   
and whether there are evolutionary links between blue GC candidates
in E+A's and the red metal-rich GC's around  passive ellipticals. 
We concentrate in particular on the 2D dynamical and spectroscopic distributions 
and the GC properties of our model merger/interaction remnant galaxies
{\it when they have an H$\delta$ equivalent width, EW(H$\delta$), larger
than 6\AA}, consistent with stringent criterion adopted 
by Z96 in their E+A selection. We here stress that when the simulated
model shows strong EW(H$\delta$), other Balmer absorption lines also
show large values: The simulated models with EW(H$\delta$) larger than
6\AA \,  are consistent with Z96's selection criterion. 
 
The star formation histories, which determine the final spectrophotometric
properties of the remnant interacting/merging galaxies,
differ quite significantly in our simulations, being dependent on the
gas mass fraction ($f_{\rm g}$), the bulge-mass fraction ($f_{\rm b}$),
and the mass ratio of the merging two disks ($m_{2}$). Figure~1 shows
an example of this dependence. The structural and kinematical properties 
of the simulated E+A's are also diverse, depending mainly on $m_{2}$ and 
the orbital configurations.
We therefore describe the results of our most representative models which 
show the typical and/or most interesting behavior in E+A formation and evolution.
We first describe the physical properties of the simulated E+A
in the fiducial merger model M5, and then discuss their parameter dependences
based on the results of other representative merger models.
The morphological properties seen from the orbital plane of the galaxy 
interaction/merger (e.g., the $x$-$z$ plane) are shown in Figure~2 
for the representative models, M1, M5, U1, and T1.
We mainly describe the results of these representative models.

We will explain separately the results of the merger models and
the tidal interaction models, in order to contrast the differences
in properties between the two. In the following sections, 
the time $T$ represents the time that has elapsed since the simulation starts.

\begin{figure}
\psfig{file=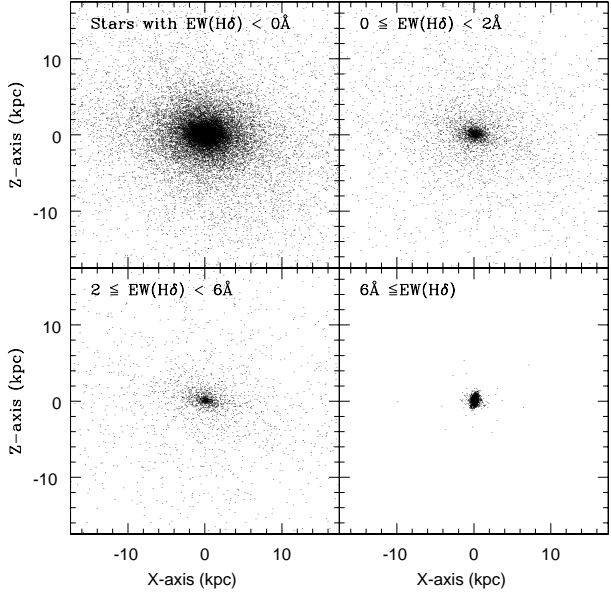,width=8.cm}
\caption{ 
Mass distributions projected onto the $x$-$z$ plane
in the fiducial model at $T$ = 2.8 Gyr
for stellar populations with different EW(H$\delta$):
For EW(H$\delta$) $<$ 0 \AA \, (upper left),
0 $\le$ EW(H$\delta$) $<$ 2 \AA \, (upper right),
2 $\le$ EW(H$\delta$) $<$ 6 \AA \, (lower left),
and 6 \AA \, $<$ EW(H$\delta$) (lower left).
}
\label{Figure. 3}
\end{figure}

\begin{figure}
\psfig{file=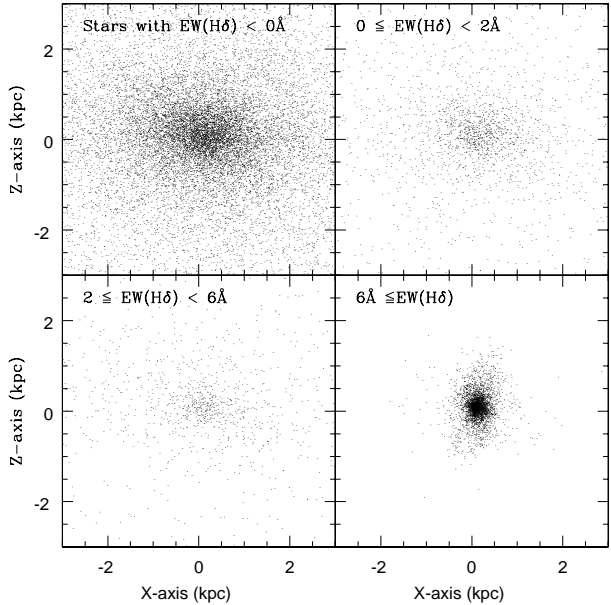,width=8.cm}
\caption{ 
The same as Fig. 3 but for the central 3 kpc. 
}
\label{Figure. 4}
\end{figure}

\begin{figure}
\psfig{file=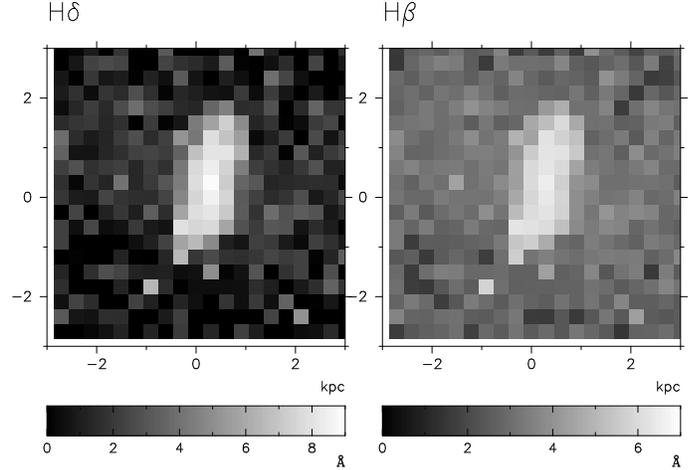,width=9.cm}
\caption{ 
The two-dimensional (2D) distributions of EW(H$\delta$) ({\it left}) and 
EW(H$\beta$) ({\it right}) projected onto the $x$-$z$ plane in the fiducial model 
at $T=2.8$\,Gyr. 
The abscissa and the ordinate represent the $x$-axis and the $z$-axis, respectively. 
Here we divide the 3\,kpc $\times$ 3\,kpc central region (shown in Fig. 4)
into 20 $\times$ 20 grid points and thereby 
estimated the SED of each grid point.
For clarity, the grid points with  EW(H$\delta$) (and EW(H$\beta$))
less than 0\AA\, are shown in the darkest color.
Note that the 2D distributions show the strong Balmer line absorption 
to be elongated along the $z$-axis.
}
\label{Figure. 5}
\end{figure}

\begin{figure}
\psfig{file=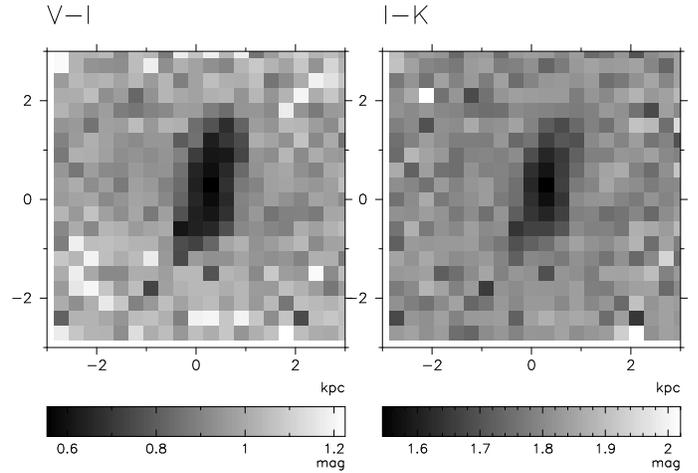,width=9.cm}
\caption{ 
The same as Fig. 5 but for the optical color $V-I$ ({\it left})
and the near-infrared color $I-K$ ({\it right}).
}
\label{Figure. 6}
\end{figure}

\begin{figure}
\psfig{file=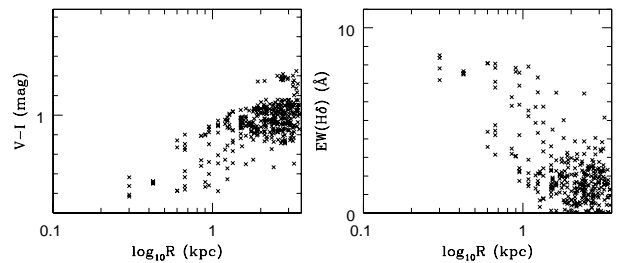,width=8.cm}
\caption{ 
Radial profiles of $V-I$  ({\it left}) and EW(H$\delta$) ({\it right}) derived
from $V-I$  and EW(H$\delta$) of the 400 grid points shown
in Figs. 5 and 6. 
}
\label{Figure. 7}
\end{figure}

\section{Results}

\subsection{The fiducial merger model}

\subsubsection{2D distributions of spectrophotometric properties}

Figure 3 summarises the global morphological properties
of the stellar populations with different EW(H$\delta$) for 
the E+A formed in the fiducial model at $T$ = 2.8\,Gyr.
In this major merger model with a highly inclined orbital 
configuration and a large pericenter distance (17.5\,kpc),
the old stars with EW(H$\delta$) $<$ 0 \AA \,
appear to have the morphology on an elliptical 
galaxy. Furthermore, there is no sign of any tidal features,
because the tidal tail formed during the merger disappeared
by the time the remnant shows strong Balmer line absorption.
The major axis of the E+A is nearly coincident with
the $x$-axis  for the old stars.

The young  stars that are formed during the merger (with 
$0\le$EW(H$\delta)<6$\AA), have more compact spatial
distributions than the old stars, although their major axes
are nearly aligned with those of the old stars.
Young stars with EW(H$\delta)>6$\AA \, have the most compact
spatial distribution, and are confined to 
the central few kiloparsecs. This reflects the fact that these stellar
populations were formed in the last major starburst during the
merger. 

Figure~4 shows that there is a clear difference in the morphological 
properties between the old stars (with EW(H$\delta)<0$\AA)
and the youngest stars (with EW(H$\delta)>6$\AA) 
within the central 3\,kpc of the E+A formed in the fiducial model.
In the central 3\,kpc, the major axis of the old stars' mass distribution 
is nearly aligned with the $x$-axis and thus
with that of the {\it global} mass distribution (shown in Fig.~1), 
whereas the major axis of the young stars 
is nearly parallel to the $z$-axis (and thus perpendicular to that 
of the mass distribution of old stars).
This misalignment in the mass distributions of the E+A results from
the dissipative formation of a thick stellar disk 
composed of stars with strong Balmer absorption lines in the E+A core. 
The more diffuse distribution of young stars (with $0\le$EW(H$\delta)<6$\AA), 
is due to the earlier formation of these stars within the merger progenitor
disks (rather than in the very center of the merger at the final phase of merging). 

Figure~5 clearly shows that the 2D distributions of EW(H$\delta$)
and EW(H$\beta$) absorption have very flattened shapes along the $z$-axis,
and the direction of elongation is coincident with the major axis of 
the mass distribution of young
stars with  EW(H$\delta)>6$\AA (shown in Fig.~4). 
It also shows that both EW(H$\delta$) and EW(H$\beta$) are larger in the inner
regions of the E+A, and therefore have a negative radial gradients.
These results are all due principally to the centralized starbursts 
during dissipative merging in this model.
The flattened  2D distribution can also be seen in the optical ($V-I$) 
and near-infrared ($I-K$) color distributions shown in Figure~6,
which confirms that the flattened shape is due to 
the flattened distribution of young stars
in the core of the E+A's.

Figure~7 shows that the radial gradient in $V-I$ color
is ``positive'' in the sense that the inner color is bluer;
this is also true for $I-K$. This figure also shows graphically, 
the negative radial gradients seen in EW(H$\delta$) and EW(H$\beta$). 
We find that the mean logarithmic gradient in $V-I$, i.e.,
$\Delta (V-I)/\Delta (\log R)$ is 0.14\,mag per dex in radius, $R$, whereas 
that in  EW(H$\delta$) ($\Delta H\delta/\Delta(\log R)$)
is $-2.6$\,\AA \, per dex in radius $R$. 
These positive color gradients and negative EW(H$\delta$) gradients in
our simulated E+A's are consistent with what is observed in {\it some} E+A's
(e.g., Yang et al. 2004).
We discuss such comparisons further in \S 4.
 
\begin{figure}
\psfig{file=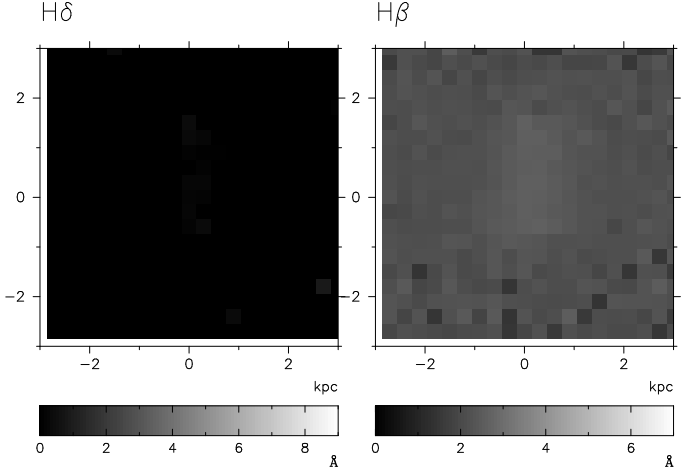,width=9.cm}
\caption{ 
The same as Fig. 5 but for $T$ = 4.5\,Gyr,
when the model shows a passive (``k''-type) spectrum. 
}
\label{Figure. 8}
\end{figure}

\begin{figure}
\psfig{file=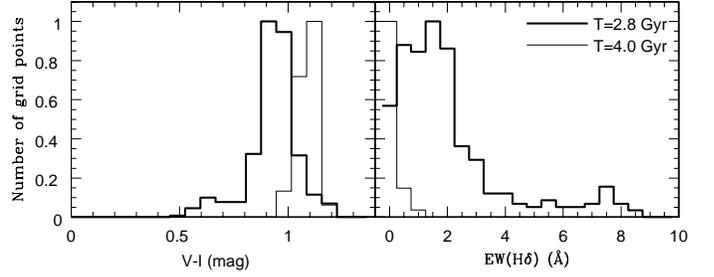,width=9.cm}
\caption{ 
Distributions in the number of grid points with a given color ({\it left})
and EW(H$\delta$) ({\it right}) for two different epochs, $T=2.8$\,Gyr ({\it thick 
solid line}) and $T=4.0$\,Gyr ({\it thin solid line}), in the fiducial model.   
The color and EW(H$\delta$) values are those at each grid point in the 
2D distributions (as shown in Figs. 5 and 6) and 
the normalized number of grid points is shown for convenience.
The spectral type of the simulated E+A evolves from ``a+k'' to ``k+a'' (to ''k'') 
during these two epochs. Note that the dispersion of the distribution
becomes very small in the k+a phase (i.e., $T=4.0$\,Gyr) both for $V-I$ and 
EW(H$\delta$).
}
\label{Figure. 9}
\end{figure}

\begin{figure}
\psfig{file=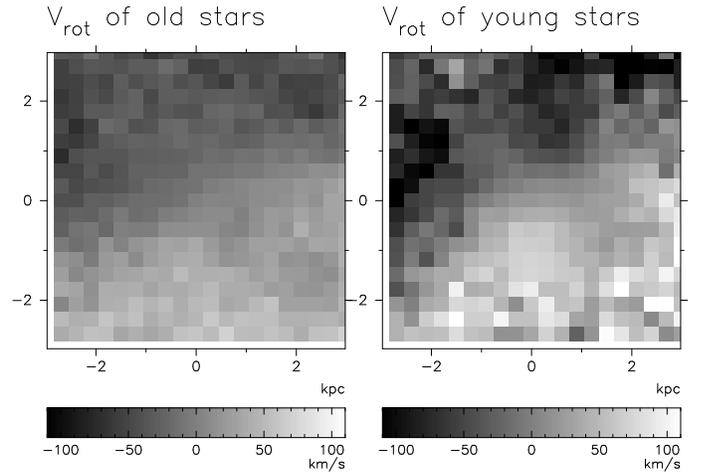,width=9.cm}
\caption{ 
The 2D distribution of line-of-sight velocities viewed from the $y$-axis
(i.e., line-of-sight velocity field projected onto the $x$-$z$ plane)
for old stars ({\it left}) and young stars ({\it right}) in the fiducial model
at $T=2.8$\,Gyr. The abscissa and the ordinate represent the $x$-axis and the 
$z$-axis, respectively. This figure enables us to confirm whether the simulated 
E+A has global rotation and which direction the ZVC extends. 
The ZVC extends from the lower left corner to the upper right corner in
this figure for old stars. 
Each frame measures 6\,kpc so that this figure can be compared with
Figs. 4, 5, and 6.
}
\label{Figure. 10}
\end{figure}

\begin{figure}
\psfig{file=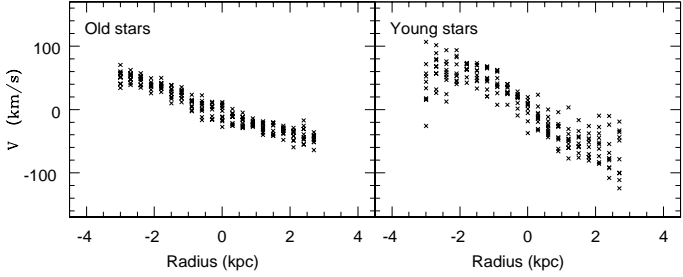,width=9.cm}
\caption{ 
Radial profiles of line-of-sight velocity along the $z$-axis
derived from grid points shown in Fig.~10 for old stars ({\it left})
and young stars ({\it right}). 
Note that the radial gradient is steeper for the young populations
than for the old ones, though the dispersion in the outer part is
large for the young ones.
}
\label{Figure. 11}
\end{figure}

\begin{figure}
\psfig{file=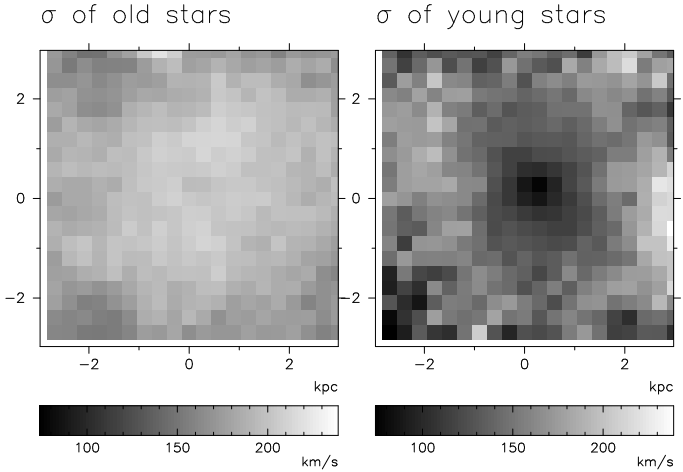,width=9.cm}
\caption{ 
The same as Fig.~10 but for the velocity dispersion.
Note that the young stars show smaller dispersion in the inner regions
of the E+A. 
}
\label{Figure. 12}
\end{figure}

\begin{figure}
\psfig{file=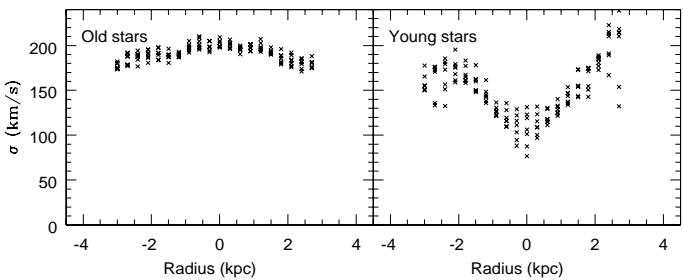,width=9.cm}
\caption{ 
The same as Fig.~11 but for the velocity dispersion.
}
\label{Figure. 13}
\end{figure}

In the E+A simulated in the fiducial model, the spectrophotometric properties
-- as delineated by the 2D distributions -- rapidly evolve with time, due 
to the fading of the young
stars. Figure~8 shows that the elongated distribution 
in Balmer line absorption seen in the strong E+A phase
($T=2.8$\,Gyr) becomes nearly invisible at $T=4.5$\,Gyr 
(i.e., 1.7\,Gyr after the strong E+A phase). This disappearance of the 
flattened shapes is also seen in the 2D $V-I$ and $I-K$ distributions.
Figure~9 describes how the 2D $V-I$ and EW(H$\delta$) distributions
evolve from the E+A phase to the passive ("k" type) phase. 
It is clear from this figure that the dispersion in the distribution
becomes significantly smaller as the E+A elliptical evolves into a passive 
system both for $V-I$ and EW(H$\delta$). The $V-I$ (EW(H$\delta$))  dispersion 
decreases from 0.12\,mag (2.2\,\AA) to 0.04\,mag
(0.8\,\AA) during the transition phase. These results suggest that the 
dispersion in the 2D spectrophotometric properties can be regarded as an 
evolutionary ``time arrow'' in going from the E+A phase to passive phase.
We thus suggest that elliptical galaxies with weaker H$\delta$ have
smaller dispersion in the 2D distributions of colors and Balmer absorption lines.

\subsubsection{Kinematics}
 
Figure~10 displays the 2D distribution of line-of-sight-velocities 
for the old and young stars for the simulated E+A of the fiducial model 
at $T=2.8$\,Gyr. It can bee seen that for the old stars, the zero velocity 
curve (ZVC) -- which 
is defined as a line connecting the points where the line-of-sight velocity 
is zero -- extends from the lower left corner to the upper right corner 
and thus is not aligned with the major axis of the mass distribution of old stars
shown in Figs. 3 and 4. The direction of the ZVC of young stars with 
EW(H$\delta)\ge 2$\,\AA, is broadly consistent with that of the ZVC of old stars,
though there is a clear difference in the velocity field along the $z$-axis
(i.e., the vertical axis) between the two components for $|x|\le 2$\,kpc.
The misalignment seen between the ZVC and the major axis of the mass
distribution in E+A's, both for old and young stars, 
clearly indicates the minor-axis of rotation of the E+A.  
 
Figure~11 shows the radial rotation profile along the minor axis 
(i.e., $z$-axis) derived from the data points shown in  Figure~10. 
Figure~10  shows that the rotation profile along the minor axis
is significantly different between the old stars and the young stars.
By applying a least-squares fit method to the grid points
with  $|x|\le 1.5$\,kpc in Figure~10, we can quantify the difference 
in the minor-axis rotation between the two. The derived rotational 
velocity, $V_{\rm rot}$, expressed as a function of  
the $z$ coordinate (in units of kpc), is given by:
\begin{equation}
V_{\rm rot} \approx -18.5 \times z -2.5 
\end{equation}
for the old stars, and 
\begin{equation}
V_{\rm rot} \approx -27.0 \times z -1.2, 
\end{equation}
for the young stars, where the velocities are in units of km\,s$^{-1}$. 
In the same way, the major axis rotation profile (along the $x$-axis)
can be estimated. The derived  $V_{\rm rot}$, expressed 
as a function of the $x$ coordinate is:
\begin{equation}
V_{\rm rot} \approx 4.4 \times x -1.9
\end{equation}
for the old stars, and 
\begin{equation}
V_{\rm rot} \approx 9.3 \times x -0.2 
\end{equation}
for the young stars.

These results clearly indicate that the young stars 
are more strongly supported by rotation
than the old stars in the model E+A. 
The origin of the steeper radial gradient of $V_{\rm rot}$ 
along the minor axis is closely associated with the formation
of the kinematically distinct core (KDC) composed mostly of 
young stars with strong  H$\delta$ absorption 
in the central few kpc of the E+A. 
Owing to the efficient gaseous dissipation during the very late phase
of galaxy merging, the central core can have a significant amount of rotation
with the angular momentum axis different from that of the major old component.
As a result of this formation of a KDC, 
the minor axis rotation can be more remarkable
in young stars than in old ones in the 2D velocity distribution.
Thus Figures 5 and 10 suggest that {\it future observations can confirm
the dissipative formation of KDCs in galaxy mergers  by investigating the
2D distribution of line-of-sight velocities of young stellar populations
in E+A's.} 

Figure~12 shows the 2D distribution of line-of-sight velocity dispersion ($\sigma$)
both for old and young stars in the simulated E+A. Although the dispersion 
field appears to be somewhat irregular, in particular, for the young stars,
it shows that the central dispersion is higher for the old stars than the 
young stars and the radial gradient of the dispersion is ``positive'' for young
stars in the sense that the inner velocity dispersion is lower in
the inner regions than in the outer regions for the central few kpc of the E+A.
These two results do not depend on parameters of the present merger models,  
so that they can be considered to be generic characteristics of the simulated E+A's.

Figure~13 shows that the radial gradient of velocity dispersion is steeper
for the young stars the older ones. If we assume $\sigma$ $\propto$ $R$, 
where $R$ is the distance of each grid point in Figure~12 from the center of the E+A, 
then we can roughly quantify the difference in the radial gradient
between the two components based on the least-squares fit method.
The variation of the derived velocity dispersion, $\sigma$, with 
radius, $R$,  for $R$ $\le$ 3 kpc is described by:  
\begin{equation}
\sigma \approx -11.5 \times R + 210.0 
\end{equation}
for the old stars, and 
\begin{equation}
\sigma \approx 8.3 \times R + 130.0 
\end{equation}
for the young stars.
The reason for the young stars apparently having such a small $\sigma$ 
slope is that the dispersion in $\sigma$ is 
large both at larger radii ($2-3$ kpc) and at the centrer
(See Fig.~13): The least-squares fit to data points with the 
projected distance ($R$) less than 3 kpc give a small $\sigma$ slope
owing to this large dispersion.
It should be stressed here that if the $\sigma$ slope in young stars
is estimated for data points with $R$ less than
1.5 kpc, the slope becomes significantly steeper 
($\sigma \propto 35 \times R$). 
We discuss the consistency of these results with recent observations by
Norton et al. (2001) later in \S 4.

It should be stressed here that the positive radial gradient in velocity 
dispersion we see for the young stars, is consistent with the radial velocity 
dispersion profiles observed in bright nucleated dwarf elliptical galaxies (dE,Ns)
in the Virgo cluster (Geha et al. 2002). This consistency implies that 
{\it dissipative processes associated with nuclear starbursts} can be responsible 
for the formation of the lower velocity dispersion nuclei in dE,N's.
Given the fact that the velocity dispersion of the young stars in the E+A galaxies, 
EA5, EA7, and EA15, observed by Norton et al. (2001) also decreases inwardly, 
there could be  an  evolutionary link between such low luminosity E+A's and dE,Ns.

\begin{figure}
\psfig{file=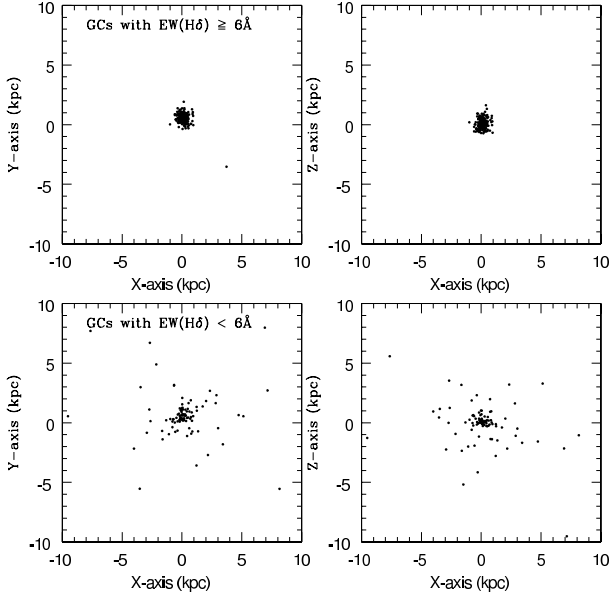,width=8.cm}
\caption{ 
The distributions of young GC's with EW(H$\delta)\ge 6$\,\AA \, 
({\it upper two panels}), and with EW(H$\delta)<6$\,\AA \, 
({\it lower two panels}), projected onto the $x$-$y$ plane 
({\it left two panels}) and onto the $x$-$z$ plane ({\it right two panels})
in the fiducial model with $T=2.8$\,Gyr.
}
\label{Figure. 14}
\end{figure}

\begin{figure}
\psfig{file=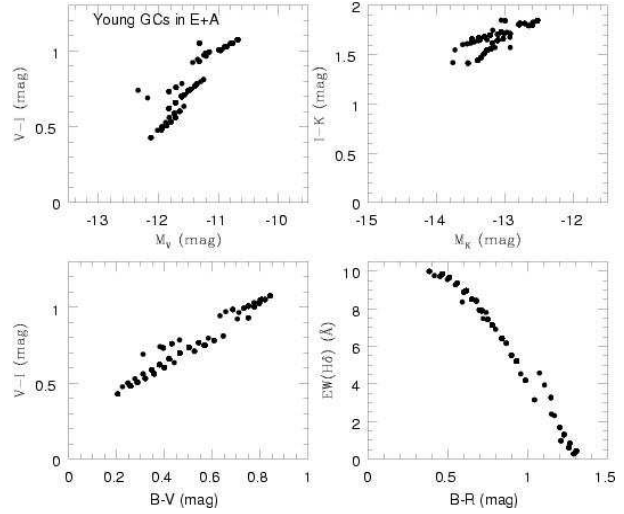,width=8.cm}
\caption{
The distributions of young GC's 
on the $(V-I)$-$M_{\rm V}$ plane ({\it upper left}),
the $(I-K)$-$M_{\rm K}$ plane  ({\it upper right}),
the $(B-V)$-$(V-I)$ plane  ({\it lower left}),
and the $(B-R)$-EW(H$\delta$) plane  ({\it lower right}), 
for the fiducial model with $T=2.8$\,Gyr.
}
\label{Figure. 15}
\end{figure}

\subsubsection{GC properties}

Figure~14 shows the spatial distributions of young GC's formed
from high-speed cloud-cloud collisions with small impact parameters
in the fiducial model. 
The mass fraction of GC's to field stars in this model is 7\%, 
that is a factor of $\sim$8 larger than that in the isolated model I1 (1\%).
This suggests that GC formation is more strongly enhanced than
field star formation in dissipative major merging. 
Nearly all (100 \%)  of young GC's with H$\delta\ge 6$\,\AA\, are within
the central 3\,kpc (i.e., the projected distance of $R$)
less than 3 kpc) of the E+A whereas about 70\% 
of young GCs with  H$\delta <6$\,\AA\ are within
the central 3\,kpc, which indicates an age-dependent spatial distribution
for the young GC's in the E+A (i.e., a more compact distribution for the younger 
GC's). This result implies that the vast majority of H$\delta$ strong
GC's will be very difficult to identify, observationally, as GC's associated 
with the E+A, due to their close proximity to their host's center. 
Only the outer young GC's with relatively small EW(H$\delta$) will be detected, 
appearing as relatively bright blue compact sources. Most of these young GC's 
have a metallicity that is higher than solar, and hence will evolve into red, 
metal-rich GC's a few Gyr later.

Figure~15 shows that young GC's in the E+A have relatively bright luminosities 
with $M_{\rm V}\ge$-10\,mag and $M_{\rm K}\ge$ -12\,mag. This is 
because of the young age of the GC's and the relatively 
high value (1.0 ${\rm M}_{\odot}$) that we have adopted for the 
lower-mass cut off of the IMF ($m_{\rm L}$).
It should be noted here that if we adopt the lower value of 0.1 ${\rm M}_{\odot}$
for $m_{\rm L}$, the above absolute magnitudes of the GC's become significantly 
(several magnitude) fainter. Since young GCS are all assumed to have the same
mass in the present study, the older (fainter) GC's will have redder colors,
as shown in Figure~15. The dispersion in the optical colors, $V-I$, of the GC's 
is large, not because the GC's have diverse metallicities but because there
are age differences between them. 
The GC's are actually distributed with a small dispersion along a line 
in the $(B-V)$-$(V-I)$ plane, due to their narrow range in metallicity.

The locations of the GC's on the $(B-R)$-EW(H$\delta$) plane in Figure~15
clearly demonstrates that they are on the ``poststarburst evolutionary path''
first identified by Couch \& Sharples (1987), where they evolve from blue 
H$\delta$ strong objects to red H$\delta$ weak objects. Hence 
blue H$\delta$ strong and red  H$\delta$  weak GC's appear to {\it coexist}
in E+A's formed via major merging with strong starbursts. 
This results also implies that E+A's formed via other physical mechanisms, 
such as truncation of star formation without a starburst, do not
show such coexistence in their GC systems. 
Red GC's in passive ellipticals can be either GC's that were initially
associated with the merger progenitor spiral or those that were formed during
dissipative merging. Bright,  blue GC's with strong H$\delta$ absorption lines 
in E+A's are unambiguously {\it young GC's formed during dissipative merging}. 
We therefore suggest that if the derived age-dependent spatial distribution  
and spectrophotometric properties of young GC's in the simulated E+A's
can be compared with future observations, such theories on the formation 
of GC's during mergers can be tested in a more stringent way.

\begin{figure}
\psfig{file=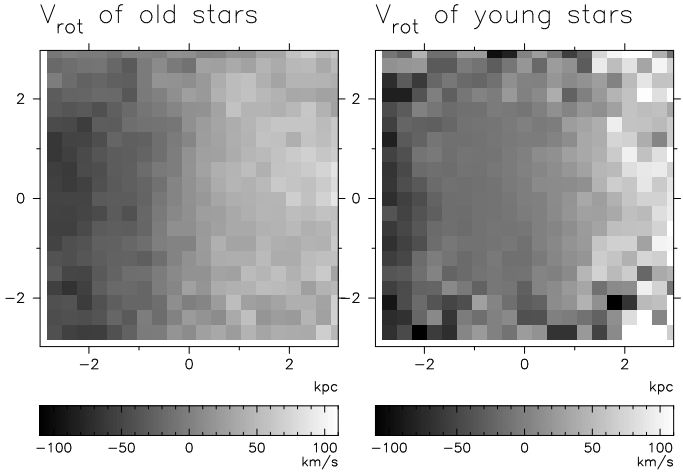,width=9.cm}
\caption{ 
The same as Fig.~10 but for the model M1 at $T=2.8$\,Gyr.
}
\label{Figure. 16}
\end{figure}

\begin{figure}
\psfig{file=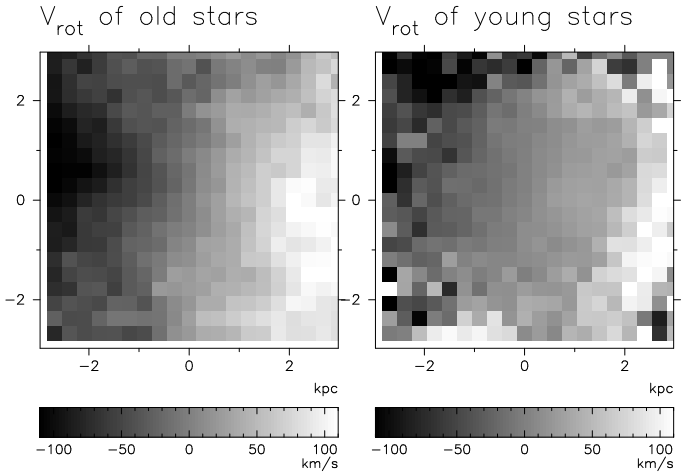,width=9.cm}
\caption{ 
The same as Fig.~10 but for the model U1 at $T=2.8$\,Gyr.
}
\label{Figure. 17}
\end{figure}

\subsection{Parameter dependences of merger models}

\subsubsection{Generic results}

Before discussing the dependencies our merger model results have on the
various physical parameters, it is useful to summarise the more generic 
results of our study. These are as follows:

(i)\,E+A's formed by galaxy merging exhibit positive color gradients 
and negative EW(H$\delta$) gradients, due to the poststarburst populations
that reside within a few kpc of their centers. By mass, the fraction of
all young stars with EW(H$\delta)\ge 2$\,\AA\ 
among all new stars  can be as high as $\sim$60\% 
in the central 3\,kpc of these E+A's.
The mass fraction of  young stars with EW(H$\delta)\ge 2$\,\AA\ 
among {\it all stars} (i.e., old + new stars) can become as high
as 0.2 in the central 3\,kpc. 
This fraction depends on the gas mass ratio of $f_{\rm g}$
in such a way that the fraction is larger for larger $f_{\rm g}$
(e.g., 0.16 for $f_{\rm g}$ = 0.1 and 0.20 for $f_{\rm g}$ = 0.5).
The fractional light from these poststarburst populations
in the standard  model for $B-$band is $\sim$ 60 \% for   
$M_L=0.1\,{\rm M}_{\odot}$ and $\sim$ 100 \% for 
1.0\,${\rm M}_{\odot}$ at the strong poststarburst epoch:
The SEDs of E+A's can be largely  determined by young stars.

(ii)\,These young stars (with EW(H$\delta)\ge2$\,\AA) show more 
rapid rotation and a smaller central velocity dispersion than the old stars 
in the central 3\,kpc of E+A's. The radial gradient in the velocity dispersion
of these stars is positive in the sense that the velocity dispersion
is smaller in the inner regions of an E+A.

(iii)\,E+A's formed in this way also have young, H$\delta$-strong, metal-rich 
(more than the solar 
metallicity) and blue GC's, most of which are located  in their central regions.
The GC formation efficiency increases dramatically (by a factor of 10) during
dissipative merging, so that E+A's end up with a larger GC specific frequency 
(i.e., number of GCs per unit luminosity) compared to their progenitor disk galaxies.
Number fraction  of GC particles  to all new stellar ones
is $\sim$ 1 \% for the isolated disk
models, I1, I2 and I3 and $\sim$ 7 \% for the major merger model M1.

\subsubsection{Orbital configurations}

Another key feature of our models is that they show the kinematical properties 
of the old and young stars in E+A's depend strongly on the orbital configurations 
of the major merger event responsible for their formation. This dependency can be
summarised as follows:

(i)\,Some E+A ellipticals show a clear sign of major-axis rotation, 
yet no sign of minor-axis rotation in the 2D distribution of line-of-sight
velocities for both old and young stars. Figure~16 shows an example
of this for the major merger model M1 in which an E+A elliptical is formed
via a nearly ``prograde-prograde'' merger, with the orbital spin axis nearly
parallel to the intrinsic spin axes of the two merger progenitor disks.
The rotational velocity, $V_{\rm rot}$ (km s$^{-1}$), expressed 
as a function of the $x$ coordinate (kpc), that is 
derived for the 2D velocity distribution of the young stars in this model 
for $|x|$ $\le$  3kpc is:
\begin{equation}
V_{\rm rot} \approx 20.1 \times x +13.9 
\end{equation}
for old stars and 
\begin{equation}
V_{\rm rot} \approx 21.7 \times x +13.0. 
\end{equation}
The major-axis rotation is much less remarkable for the model M10
in which an E+A is formed via a ``retrograde-retrograde'' merger.

(ii)\,Quite flattened EW(H$\delta$), EW(H$\beta$), $V-I$, and $I-K$ distributions 
are formed in the central few kpc of the E+A if the major merger responsible 
has a highly inclined orbital configuration and at least one
disk orbiting in a retrograde manner (i.e., the intrinsic spin axis
is anti-parallel to its orbit with respect to the mass center of the merger).
The origin of the flattened distributions is due essentially to
the formation of KDCs composed of young, H$\delta$-strong stars.  
Due to the presence of young KDCs, E+A ellipticals can show 
a significant difference in the 2D line-of-sight velocity and velocity dispersion 
between their old and young stars.

\begin{figure}
\psfig{file=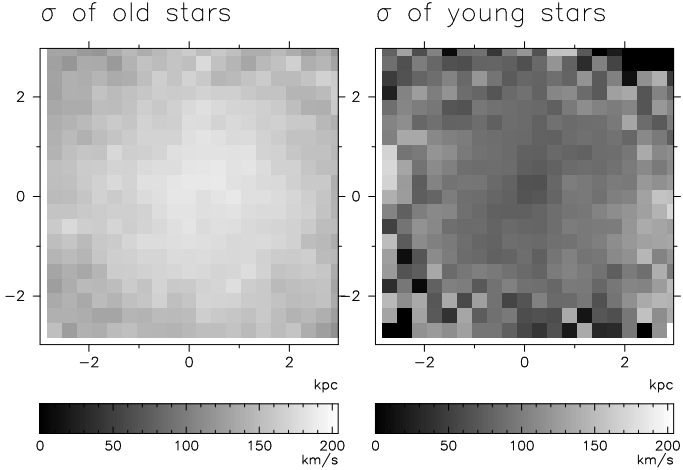,width=9.cm}
\caption{ 
The same as Fig.~11 but for the model U1.
}
\label{Figure. 18}
\end{figure}

\begin{figure}
\psfig{file=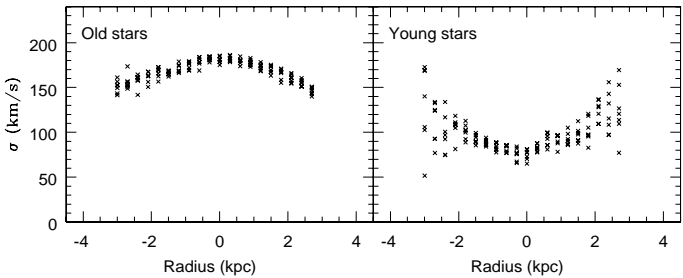,width=9.cm}
\caption{ 
The same as Fig.~13 but for the model U1.
}
\label{Figure. 19}
\end{figure}
 
\subsubsection{Mass ratio, $m_2$}

The dependences of E+A properties on the merging galaxy mass ratio, $m_2$, 
can be described as follows:

(i)\,E+A's formed by unequal-mass mergers are highly likely to show rapid rotation
along their major-axis, irrespective of the orbital configuration of the merger, 
if they are viewed edge-on. Figure~17 shows the 2D distributions of the line-of-sight 
velocities for old and young stars in the model U1, where more rapid rotation is
seen than in the M1 model shown in Figure~16. This is due to the fact that
the disk of the larger of the two progenitor spirals is not completely destroyed.

(ii)\,The positive radial gradient in the velocity dispersion
of the young stars in E+A's is not as pronounced in models with smaller $m_2$. 
This is because there is less gaseous dissipation in a  
weaker starburst and thus the young stars with relatively cold kinematics
are less centrally concentrated in these models with smaller $m_2$.
Figures~19 and 20 show the 2D distributions of velocity dispersion
and the radial profile of velocity dispersion, respectively,
for old and young stars in the model U1. It is clear from these figures
that even the unequal-mass model shows a difference in kinematics between 
the old and young stars. The variation in velocity dispersion
$\sigma$ (km s$^{-1}$) 
with  radius, $R$ (kpc) seen in these figures 
for $R$ $\le$ 3 kpc can be described as:  
\begin{equation}
\sigma \approx -15.1 \times R + 193.3
\end{equation}
for the old stars, and 
\begin{equation}
\sigma \approx 0.2 \times R + 90.9 
\end{equation}
for the young stars.

(iii)\,The mass ratio of young stars with EW(H$\delta)\ge 2$\,\AA\ 
to all new (young) stars during the post-starburst phase, is significantly 
smaller for models with smaller $m_2$ ($\le 0.1$). This is 
because the secondary starburst is much weaker in such models, 
compared to the major or unequal-mass merger models. 
For example,  the mass ratio of the strong-H$\delta$ young stars is
only 13\% in the model U2 with $m_2$ = 0.1, and  58\% for the fiducial model. 
This suggests that minor mergers with $m_2$ less than 0.1 are highly unlikely
to become E+A's after their secondary starbursts, due to the very small fraction of
A-type stars. The present models with $m_2$ = 0.1 show EW(H$\delta$) of 
$\sim$2\,\AA\ and EW(H$\beta$) of $\sim$3.5\,\AA\ in their poststarburst phases.
Therefore, the observed disky E+A's with strong Balmer absorption
lines such as those observed in Zabludoff et al. (1996) are not likely to be formed from
minor merging.

(iv)\,The formation efficiency of young GC's depends on $m_2$ in such a way
that a larger number of GC's are formed during merging in the models with
larger  $m_2$. Therefore, E+A's with flattened (S0-like) morphologies formed via
unequal-mass mergers show a smaller number of young GC's than those with elliptical 
morphologies formed via major mergers. Therefore, the observed diversity in 
the number of young GC candidates in E+A's (Yang et al. 2004) could be due to this 
diversity of $m_2$ in galaxy merging.
Flattened E+A's are likely to show flattened spatial distributions of young GCs.

\subsubsection{Miscellaneous}

Finally, we note a number of other parameter dependences that our models
have revealed, in particular ones involving $f_{\rm g}$, $f_{\rm b}$,
and the IMF:

(i)\,It is possible that two moderately strong starbursts, separated by a time
interval of $\sim$1\,Gyr, can occur during major merging
if the progenitor spirals have either no bulge ($f_{\rm b}$ =0.0)
or only a very small bulge ($f_{\rm b}$ $<$ 0.1). The first and stronger of 
these starbursts occurs as a result of the formation of a stellar bar, 
which in turn causes an inflow of gas into the center of the disk
(e.g., in the bulgeless model M9). After this first starburst phase,
the two galaxies are still separated yet at least one of them has a strong poststarburst
population. The second weaker starburst then occurs
when the two disks finally merge to form an elliptical.
Due to the rapid consumption of gas in the first starburst, 
the final E+A elliptical will, after the second starburst, have a smaller
fraction of A-type stars (39\% in the M9 model, compared to 58\% in the fiducial 
model) and thus have weaker H$\delta$ absorption (EW(H$\delta)=5.3$\,\AA\ in the 
M9 model). In the interval between the two starbursts, such a merging system would 
appear as an interacting pair with an E+A spectrum, providing the star formation rate 
is insignificant.

(ii)\,The Kinematical differences between the old and young stars are larger
in the models with larger $f_{\rm g}$ (i.e., larger gas mass fraction),
essentially because a larger amount of random kinematical energy 
in the gas can be lost during merging, due to the more efficient gaseous dissipation 
in these models. The equivalent width of H$\delta$ during the E+A phases does not 
depend strongly on $f_{\rm g}$ for a given IMF, as long as the merging galaxies 
are sufficiently gas-rich ($f_{\rm g}$ $\ge$ 0.1).

(iii)\,The equivalent width of H$\delta$ depends strongly on the IMF of the 
starburst, in the sense that EW(H$\delta$) is larger for models with larger $m_{\rm L}$.
For example, the simulated E+A in the M1 model with $f_{\rm g}$ = 0.1 has 
EW(H$\delta)=7.4$\,\AA\ for $M_{\rm L}=1\,{\rm M}_{\odot}$ 
and EW(H$\delta$)=1.1\,\AA \, for $M_{\rm L}=0.1\,{\rm M}_{\odot}$, 
whereas the E+A in the M7 model with $f_{\rm g}$ = 0.5 has  
EW(H$\delta)=6.9$\,\AA \,  for $M_{\rm L}=1\,{\rm M}_{\odot}$ 
and EW(H$\delta)=5.6$\,\AA  \, for $M_{\rm L}=0.1\,{\rm M}_{\odot}$.
In contrast, EW(H$\beta$) does not depend that strongly on the IMF.

\begin{figure}
\psfig{file=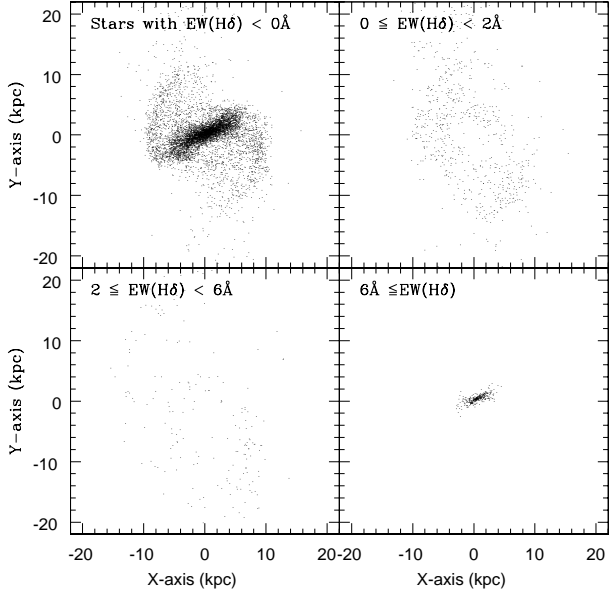,width=8.cm}
\caption{ 
The same as Fig.~3 but for the model T1 projected onto the $x$-$y$ plane.
}
\label{Figure. 20}
\end{figure}

\begin{figure}
\psfig{file=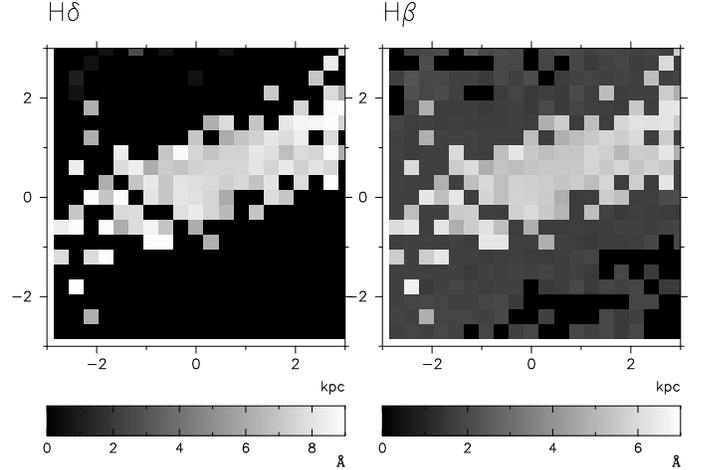,width=9.cm}
\caption{ 
The same as Fig.~5 but for the model T1 projected onto the $x$-$y$ plane.
The abscissa and the ordinate represent the $x$-axis and the $y$-axis, respectively.
}
\label{Figure. 21}
\end{figure}

\begin{figure}
\psfig{file=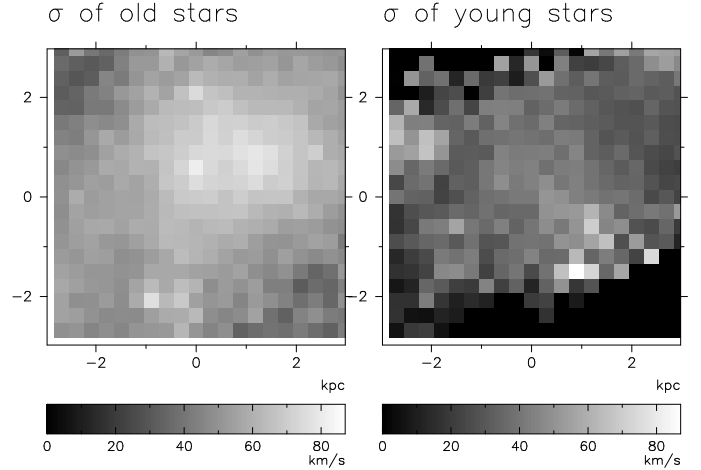,width=9.cm}
\caption{ 
The same as Fig.~12 but for the model T1 projected onto the $x$-$y$ plane.
The abscissa and the ordinate represent the $x$-axis and the $y$-axis, respectively.
The grid points with no stars  
are shown in the darkest color (i.e., $\sigma =0$\,km\,s$^{-1}$).
}
\label{Figure. 22}
\end{figure}

\subsection{Tidal interaction}

In the tidal interaction model, a strong starburst will occur 
if the two disk galaxies are of comparable masses and they interact in a prograde manner 
i.e., if the intrinsic spin axis of the disk is parallel to the orbital 
axis. This is due to the formation of a stellar bar that drives gas into the 
central regions. Figure~20 shows the mass distributions of stars with different 
EW(H$\delta$) in the tidal interaction model T1, which develops a central stellar bar
after the starburst. The mass distribution clearly depends on the EW(H$\delta$) of 
the stars: (i)\,the young stars (with EW(H$\delta)\ge$6\,\AA; {\it bottom 
right-hand panel} of Fig.~20) delineate a very compact bar in the center of the 
model, (ii)\,the old stars (with EW(H$\delta)<0$\AA; {\it top left-hand panel}
of Fig.~20) delineate a global ($\sim$10\,kpc scale) bar with two spiral arms,
and (iii)\,the intermediate age stars (with $0\le$ EW(H$\delta)<2$\,\AA; {\it top
right-hand} panel in Fig.~20) appear to have a ring-like morphology.

Because of the elongated distribution of the H$\delta$-strong stars,
the 2D distributions of EW(H$\delta$) and EW(H$\beta$) are also very
flattened. Figure~21 clearly indicates that the radial gradients 
of EW(H$\delta$) and EW(H$\beta$) along the major axis of the flattened
mass distribution of the young stars (with EW(H$\delta)\ge 6$\,\AA) 
is smaller than those perpendicular to the major axis,
though the radial gradients are essentially negative 
(i.e., large in the inner regions). The $V-I$ and $I-K$ colors also have
2D distributions with the same shapes and are also seen to have 
positive color gradients. Thus the simulated E+A has positive color gradients 
and negative gradients in EW(H$\delta$).

Figure~22 shows that the vertical velocity dispersion of the old stars 
is significantly larger than that of the young stars  
in the 2D distribution of velocity dispersion for the T1 model. 
This is because old stars are dynamically heated during tidal interaction
whereas the gas, from which young stars form, dissipates its random kinematical
energy during the interaction.
The mean dispersion is 55.5\,km\,s$^{-1}$ for the old stars and 28.8\,km\,s$^{-1}$
for the young stars. The radial dependence of the 
vertical velocity dispersion, $\sigma$ (km s$^{-1}$),  
can be estimated from the 2D distributions shown in Figure~22
and expressed as a function of $R$ (radius from the center of the E+A)
for $R$ $\le$ 3 kpc:
\begin{equation}
\sigma \approx -9.0 \times R + 76.2 
\end{equation}
for the old stars, and 
\begin{equation}
\sigma \approx -11.1 \times R + 54.3 
\end{equation}
for the young stars.
Thus the radial gradient in velocity dispersion is
steeper for the young stars than the old stars.

\begin{figure}
\psfig{file=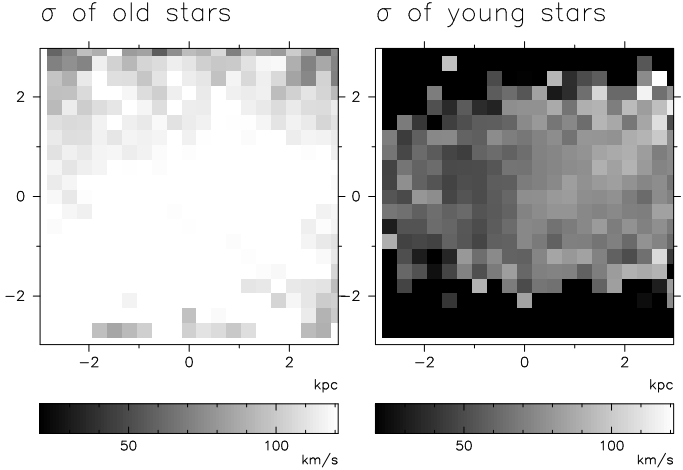,width=9.cm}
\caption{ 
The same as Fig.~12 but for the tidal interaction model T1, projected onto the 
$x$-$z$ plane.
The abscissa and the ordinate represent the $x$-axis and the $z$-axis, respectively.
In order to more clearly see the kinematical difference
between the young and old stars, the grid points with $\sigma<20$\,km\,s$^{-1}$ 
(most of which are those where no young stars
can be found) are  shown in the darkest color.
}
\label{Figure. 23}
\end{figure}

\begin{figure}
\psfig{file=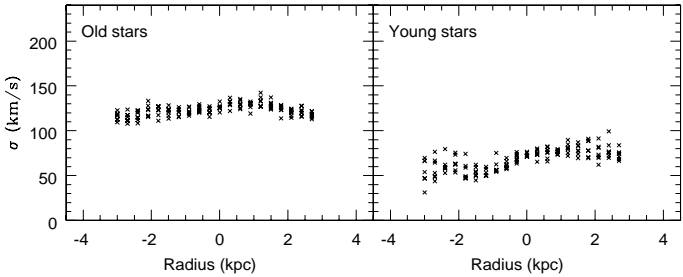,width=9.cm}
\caption{ 
The same as Fig.~13 but for the tidal interaction model T1.
The radial profile is derived for the 2D distribution shown in Fig.~23.
}
\label{Figure. 24}
\end{figure}

Figures~23 and 24 reveal quite clear differences in the 2D velocity dispersion field 
(when viewed edge-on, i.e., in the orbital plane of the interacting galaxy)  
between the young and old stars.
Overall, the old stars show a higher velocity dispersion than the young stars
within the central 3\,kpc of the simulated E+A, and a shallower radial gradient
in the velocity dispersion compared with the major merger models.
The young stars show a {\it monotonously increasing} radial profile 
in velocity dispersion over the range -3\,kpc$\le x\le 3$\,kpc
along the major axis (i.e., $x$-axis) of the mass distribution.
This non-axisymmetric profile of velocity dispersion might be an indication
that the young stars are in a non-equilibrium dynamical state in the E+A phase.
Such a non-axisymmetric profile in velocity dispersion  
cannot be seen in the 2D velocity dispersion fields viewed from
the $x$-axis and the $z$-axis, which implies that streaming of 
the gas (and young stars) along the stellar bar is partly responsible 
for the peculiar profile.  

Formation of stellar bars is essential for triggering strong starbursts
and thus for producing young poststarburst populations responsible for 
the E+A spectral signature. Therefore, the models in which bar formation 
is strongly suppressed are less likely to become E+A's. Amongst the tidal 
interaction models run for this study, T4 (with the disk orbiting in a 
retrograde manner) and T5 (with the big bulge, $f_{\rm b}$ = 1.0) 
are less likely to become E+A's, due to the weak starbursts they experience
during the tidal interaction. These results suggest that if E+A's are formed 
from tidal interactions between gas-rich spirals, the E+A's are more likely 
to be morphologically classified as barred disk galaxies, and more specifically 
as barred S0s due to the poor visibility of the spiral arms. 
Furthermore, a strong starburst associated with the formation of a stellar bar 
does not occur in the T6 model (where $m_{2}$ is small, being only 0.1),
because the tidal perturbation is not strong enough to form a bar. 
Only interaction models with a large $m_2$ ($>0.3$) show E+A spectra.
This result provides a clue to the origin of the observed low luminosity disky 
E+A's, because lower luminosity spirals are more likely to interact with  
galaxies more massive than themselves. 
Young GC's are also formed in the interaction models, 
and have a 'disky' spatial distribution, quite
different to that of the merger models described earlier.

\section{Discussion}

\subsection{Kinematical differences between old and young stars
in E+A's}

Norton et al. (2001) were the first to measure the spatially resolved kinematic 
properties of young and old stellar populations in E+A's and they
concluded that (i)\,both rapidly rotating and slowly rotating or 
non-rotating systems can be found in the E+A phase, (ii)\,most E+A 
galaxies (70\%) show no evidence for rotation,
(iii)\,the rotation is generally seen in both the old and young
stellar populations in E+A galaxies, (iv)\,the velocity dispersion ranges 
from $\sim$30\,km\,s$^{-1}$ to $\sim$200\,km\,s$^{-1}$, 
and (v)\,the young populations have {\it on average} a higher velocity 
dispersion than the old populations. In addition, Norton et al's measurements of
the radial profiles in velocity dispersion for their E+A galaxy sample (see their
Fig.~4), leads us to conclude that (vi)\,about half of the galaxies show inwardly 
decreasing dispersion profiles for the young stellar populations, 
with some at least showing the young stars to have a smaller central velocity 
dispersion than the old stars (e.g., EA5, EA7, and EA15). 

Conclusions (i)--(iii) above are consistent with our merger
models in which E+A's show a diversity in the degree of rotation  
of both the old and young stellar populations. The fourth conclusion 
can also be understood in terms of differences in the masses
of the merger progenitor spirals, because more massive E+A's with 
a larger velocity dispersion are formed in mergers between more 
massive spirals with larger rotational velocities. However, conclusion 
(v) appears to be inconsistent with the present models in which the young stars
in nearly all of the simulated E+A's show a smaller central velocity dispersion 
than the old stars. This inconsistency might be due to the different methods
that we and Norton et al. have employed to derive the velocity dispersion of the 
young stars. However, it could also be due to the fact 
that the interstellar gas in the simulations dissipates away much more random
kinematic energy during galaxy merging than actually occurs.
The dynamically cold young stellar populations 
in the simulations could well be due to 'over-dissipation'
in the gas.

Gaseous dissipation, which is responsible for the rapid gas inflow
that leads to a massive starburst during merging,
is essential for E+A formation as part of this process. 
Furthermore, the fact that our galaxy merger models
are consistent with (vi) above also points to the need for efficient
gaseous dissipation, since it is needed to produce 
dynamically cold young stellar systems. 
Therefore, the inconsistency between our models
and conclusion (v) implies that the present numerical method 
for handling gaseous dissipation and star formation
does not allow a proper treatment of the interstellar gas dynamics
in galaxy mergers. 
It therefore remains unclear how the young stellar populations can have
a larger velocity dispersion than the old stellar populations, if
they are formed from dissipative processes of star formation.

Important factors other than the 'over-dissipation' of gas 
could be responsible for the larger velocity dispersion
of young stars in E+A's. For example, if a  massive black hole (MBH) 
and young A-type stars {\it coexist} in the central 10\,pc region of
an E+A, the central velocity dispersion of the young stars
will be significantly higher than that of the surrounding older stars, 
due to the deep potential well of the MBH. If this is the case, 
the observed diversity in the differences in central velocity dispersion
between old and young stars in E+A's could be ascribed to
the existence or non-existence of MBHs in the center of E+A's. 
Thus two key lines of investigation in future studies will be to 
investigate whether an improvement in our numerical modeling
of the interstellar gas leads to more consistent agreement with (v)
above, and whether central MBHs in the progenitor galaxies 
influence the radial profiles in velocity dispersion
in E+A's formed via interactions and merging.

\subsection{Origin of negative color gradients in E+A's}

The color gradients in E+A galaxies can be considered to 
be one of the key physical properties which help us 
to determine the most plausible physical mechanism(s) of 
E+A formation. This is because the spatial distribution 
of the most recent stars formed in an E+A depends strongly 
on how the star formation was truncated, and this is well
traced by the color gradients (e.g., Rose et al. 2001).
Bartholomev et al. (2001) investigated the color gradients 
in 24 E+A galaxies, as well as those in 46 other galaxies
with normal spectra. They found that, on average, the E+A's 
had more positive color gradients than the normal galaxies. 
This is consistent with a merger/interaction origin for
the E+A signature.

However, the origin of E+A galaxies with {\it negative} color 
gradients -- which are observed both in the field and in clusters 
(e.g., Bartholomew et al. 2001, Yang et al. 2004) -- remains unclear.
In particular, the central red colors in E+A's {\it with negative 
H$\delta$ gradients} (e.g., EA2 and EA4 in Yang et al. 2004) 
are not easily explained within the interaction/merger scenario 
for E+A formation. We suggest here three possible explanations
for the existence of E+A galaxies with negative color gradients:

Firstly, the redder central colors maybe due to heavier dust 
extinction associated with the central poststarburst population.
The question then is why do only a fraction of E+A's have
dusty gas surrounding their A-type stellar populations,  
and how can dusty gas still exist given that most of the residual gas 
should be blown away from the central regions of E+A's
by supernovae explosions associated with the secondary starburst? 

A second possible explanation is that the IMF during the starburst 
is truncated, in that no stars with masses smaller than $2$ ${\rm M}_{\odot}$
are formed. Hence the poststarburst populations would be dominated 
by red evolved  stars (rather than by blue, main-sequence stars), which
would give rise to the redder colors in the E+A phase (Charlot et al. 1993).
The problem with this scenario is that it is unclear under what 
physical conditions the formation of stars with masses smaller 
than $2$ ${\rm M}_{\odot}$ can be {\it preferentially} suppressed.

Another possible explanation is that E+A's with negative color gradients 
are actually not poststarburst galaxies but are starburst galaxies heavily 
obscured by dust. Although there is marginal evidence that some cluster 
galaxies with E+A spectra show radio emission that could be powered by star 
formation activity (Smail et al. 1999), radio continuum observations by 
Miller \& Owen (2001) of the field E+A's in Zabludoff et  al's (1996)  
sample yielded only 2 detections, with radio luminosities consistent with 
only a moderate level of star formation (inconsistent with vigorous starbursts 
with $10-100$ ${\rm M}_{\odot}$). Furthermore, ultra-luminous infrared galaxies, 
some of which are believed to be dusty starburst galaxies, do not have E+A
spectra, but rather ``e(a)'' spectra where [OII]$\lambda$3727 emission is
seen in addition to the strong Balmer line absorption (Poggianti \& Wu 2000).

Unfortunately, the data currently available for E+A galaxies does not allow 
us to constrain the degree of dust extinction, the exponent of the IMF, nor the
spatial distribution of dust. Therefore it is not possible at the moment
to determine which of the above three explanations are the most likely.
It should be stressed that we have assumed in this present study that
most E+A galaxies are formed from galaxy interactions and merging. 
It might be necessary to relax this assumption to the extent that 
two or more physical processes might be responsible for E+A formation.
We will discuss this possibility in our forthcoming papers.

So far we did not discuss color gradients of E+As formed by
merging {\it between an old giant elliptical and a gas-rich
spiral},  because no theoretical studies (including the present one)
have investigated spectrophotometric evolution of this type of 
galaxy mergers. Although it  is not clear this `E-Sp' merging
can create E+A's,  it is doubtlessly worthwhile for us to investigate
this possibility in our future papers, given the fact that
some Es show positive Balmer line gradients.
E+A's formed by E-Sp merging could show positive color gradients,
if poststarburst components originally in the spiral progenitors
are dispersed into the outer parts of 
the remnants (owing to the tidal destruction
of the spirals)  and thus located in the outer part of the  remnants.

\subsection{Origin of kinematically decoupled cores in elliptical galaxies}

About one third of nearby luminous elliptical galaxies are observed
to show peculiar core kinematics, such as counter-- or oddly--rotating
cores, and are thus considered to have rapidly spinning
thick disks or torus-like components (see Bender 1996 for a review).
Numerical simulations by Hernquist \& Barnes (1991) showed that ellipticals 
with counter-rotating cores can originate from {\it dissipative} major 
galaxy merging between two spirals. If E+A ellipticals are formed by 
dissipative major merging, as shown in the present study,
the implication therefore is that a significant fraction of E+A's must 
possess kinematical distinct cores (KDCs) that are very bright and 
have strong H$\delta$ absorption. Although Norton et al. (2001) have 
already investigated possible differences between the kinematics of 
the old and young stellar populations, they did not find significant 
kinematical differences between the two populations. However, their results 
are based on long-slit spectroscopy of the stellar populations in only the 
central few kiloparsecs of E+A galaxies, and it remains unclear 
whether the kinematics of this region are different to those in
the more outer regions.

The present simulations have demonstrated that an E+A galaxy formed in
a major merger will have a very flattened 2D distribution in H$\delta$, 
resulting from the central disky core composed mostly of A-type stars.
They have also shown that the 2D kinematical distributions of the young 
stars are significantly different from those of the old stars,
which reflects the fact that the young stellar population in the core 
is more strongly supported by rotation. These results suggest that 
E+A galaxies might well be the best 'signposts' of KDC formation in 
dissipative major merging. 
Future spatially resolved, integral field
unit spectroscopy of E+A galaxies with 8-10\,m class telescopes should
be capable of mapping their 2D kinematical and spectral properties
to the level required to address the origin of KDCs in ellipticals.   
If such observations also confirm the diversity in the 2D H$\delta$ 
and kinematical distributions seen in our simulations, they 
will also confirm the dissipative major merger scenario for E+A formation.

\subsection{Formation of disky E+A's}
Recent morphological studies of E+A's have revealed that they are not
all spheroidal systems, with a certain fraction having disks, and some
of these being disk-dominated (Blake et al. 2004; Tran et al. 2004).
Moreover, there seems to be environmental differences in that more 
'disky' E+A galaxies are seen in rich clusters (up to $\sim$20\% of
the E+A population; Tran et al.) than in the lower-density group and 
field environment (only a $\sim$5\% fraction; Blake et al.).
These trends might suggest that different physical mechanisms are
responsible for E+A formation in clusters and in groups/the field. 
Ram-pressure effects in clusters of galaxies are suggested 
to enhance star formation activity in gas-rich spirals
without destroying their disk components (e.g.,  Bekki \& Couch 2003, 
Milvang-Jensen et al. 2003; Bamford et al. 2004) 
and thus can be considered to contribute 
to the large fraction of disky E+A's in clusters. 
Since galaxy interactions and merging are thought to only take place
in the galaxy group and field environments, the presence of disky 
E+A's raises the question as to what kinds of galaxy interactions and 
merging lead to the formation of E+A's with this morphology.
  
Our numerical simulations have demonstrated that: (i)\,rotationally 
supported and flattened E+A's with an S0 morphology, can be formed in 
{\it some} unequal-mass mergers, and (ii)\,E+A disk galaxies with 
thick disks can be formed in galaxy interactions with special initial 
parameters (e.g.,  prograde encounters). The latter of these is always 
associated with bar formation and double spiral arms (rather than 
multiple-arm structures). These results suggest that unequal-mass mergers
are only responsible for the formation of E+A's with an S0 morphology, 
and that tidal interactions can be responsible for the formation of E+A's
with barred morphologies or those with double-arm structures.
They also suggest that field E+A's without any thick disk component
can only be formed by galaxy merging and interaction.
The present numerical simulations have also shown that
the vertical velocity dispersion of {\it old stars} in disky E+A's
can be rather high (up to 60 km s$^{-1}$) owing to the dynamically hot
thick disk. Therefore high-resolution imaging of the vertical structure 
in disky E+A's and spectroscopic determination of the velocity dispersion 
of their old stars can provide valuable information on the presence of 
dynamically hot thick disks and thus assess the viability of 
the disky E+A formation scenario described in this paper.

\section{Conclusions}

We have investigated the dynamical and spectrophotometric 
properties of E+A galaxies formed via galaxy interactions
and merging, using gas dynamical simulations combined with 
stellar population synthesis codes. Our principle results 
can be summarised as follows:

(1)\,E+A ellipticals formed by dissipative major galaxy merging 
show positive radial gradients in color (i.e., bluer colors
at their center) and negative radial gradients in Balmer absorption 
line strength, due to the larger fraction of A-type stars in their
inner regions. These color and line index gradients become shallower 
as time passes by, because of the aging of the poststarburst stellar 
population. These numerical models, however, cannot explain
the E+A galaxies that are observed to have negative radial gradients
in both color and Balmer absorption line strength. 

(2)\,The dynamical and spectroscopic properties of E+A ellipticals 
formed by dissipative major galaxy merging have 2D distributions
which can be remarkably different, depending on the orbital 
parameters of the merger. Furthermore, such differences also
exist between the old and young stars in these galaxies. For 
example, E+A ellipticals with kinematically decoupled cores (KDC's) 
have a very flattened H$\delta$ absorption distribution, with
differences in rotation and velocity dispersion between
the old and young stars. Future spatially resolved, integral 
field unit spectroscopy with 8-10m class telescopes should be
able to bring the past KDC formation sites in young E+A ellipticals  
into relief and thus provide an evidence
that KDCs can be formed from dissipative major merging. 

(3)\,The 2D distributions of colors and Balmer absorption lines
in E+A ellipticals show a larger internal dispersion compared with
those of 'passive' ellipticals with weak H$\delta$ ($\sim$0\AA).
The internal color dispersion of ellipticals formed by major
merging becomes smaller as the E+A phase passes, and therefore 
there will be an anti-correlation between the size of this
dispersion and H$\delta$ strength in the post-E+A phase. 
These results imply that the large color dispersion
in the 2D photometric properties of E+A ellipticals 
provides further evidence for the transformation from
gas-rich late-type spirals into passive ellipticals. 

(4)\,Unequal-mass galaxy merging ($m_{2}$ $\sim$ 0.3) can trigger 
central starbursts and thus can transform two gas-rich spirals into 
very flattened ellipticals or S0s with an E+A spectral signature . 
These E+A's are flattened by rotation, with a large $V/\sigma$ ($>1$),
and show positive radial gradients in color and negative radial
gradients in Balmer line absorption. These flattened E+A's could be 
the missing link between gas-rich late-type spirals and passive S0s.

(5)\,The mass fraction of young stars with H$\delta >2$\,\AA\ in a 
merger remnant strongly depends on the mass ratio ($m_{2}$) of the two 
progenitor spirals in the sense that the fraction is larger in merger 
remnants with larger $m_{2}$. This is essentially because secondary 
starbursts during merging are stronger and thus produce a larger 
fraction of A-type stars. Therefore, it is highly unlikely that minor 
mergers with $m_{2}<0.1$ are responsible for E+A formation with 
strong H$\delta$ (EW$>6$\,\AA). 

(6)\,Strong tidal interaction between galaxies can also be responsible
for E+A formation, although special orbital configurations
(i.e., prograde encounters, larger $m_{2}$, and smaller bulges) are 
required for  the tidal encounters. The E+A's formed from tidal 
interactions are likely to have an SB0 morphology, with thick disks, 
positive color gradients and negative H$\delta$ gradients. Our results 
thus suggest that some fraction of disky E+A's with SB0 morphology
may be formed via strong tidal interactions.

(7)\,E+A's formed by galaxy merging and interactions can have  young, 
bright, metal-rich, and H$\delta$-strong GC's,  mostly in the central 
few kiloparsecs, because the number of cloud-cloud collisions with 
high relative velocities and small impact parameters are dramatically 
enhanced during this process. Disky E+A's are more likely to show 
a disky distribution of young GC's and rotational kinematics in the GC
population. Furthermore, the total number of young GC's is more likely 
to be larger in E+A ellipticals than in disky E+A's for a given luminosity,
because their formation efficiency is higher in major mergers than
in minor mergers and tidal interactions.

(8)\,The present models of E+A formation have difficulties in
explaining: (i)\,E+A's showing both negative radial  gradients in color
and EW(H$\delta$), and (ii)\,E+A's where the central velocity dispersion 
of the young stars is larger than for the old stars. These difficulties 
suggest that more sophisticated numerical methods for star formation and 
interstellar dynamics are required to reproduce self-consistently the 
diversity in the radial gradients of dynamical and spectrophotometric 
properties of the central poststarburst populations in E+A's.

\section{Acknowledgment}
We are  grateful to the referee Alfonso Arag\'on-Salamanca 
for valuable comments,
which contribute to improve the present paper.
KB and WJC acknowledge the financial support of the Australian Research Council
throughout the course of this work.
The numerical simulations reported here were carried out on GRAPE
systems kindly made available by the Astronomical Data Analysis
Center (ADAC) at National Astronomical Observatory of Japan (NAOJ).\\


\end{document}